\documentclass[superscriptaddress,showkeys,showpacs,preprint,pre]{revtex4}

\usepackage{amssymb}
\usepackage{amsmath}
\usepackage{multirow}
\usepackage{hhline}
\usepackage[dvips]{epsfig}
\usepackage{color}

\definecolor{red}{rgb}{1,0,0}

\begin{document}

\title{Strictly and asymptotically scale-invariant probabilistic models of $N$
  correlated binary random variables having {\em q}--Gaussians as $N\to \infty$ limiting distributions}
\author{A. Rodr\'{\i}guez}
\affiliation{Dpto. de Matem\'{a}tica Aplicada y Estad\'{\i}stica,
        Universidad Polit\'{e}cnica de Madrid, Pza. Cardenal Cisneros s/n, 28040 Madrid, Spain}
\author{V. Schw\"ammle}
\affiliation{Centro Brasileiro de Pesquisas F\'{\i}sicas, Rua Xavier Sigaud 150, 22290-180 Rio de Janeiro, Brazil}
\author{C. Tsallis}
\affiliation{Centro Brasileiro de Pesquisas F\'{\i}sicas, Rua Xavier Sigaud 150, 22290-180 Rio de Janeiro, Brazil}
\affiliation{Santa Fe Institute, 1399 Hyde Park Road, Santa Fe, New Mexico 87501, USA}

\begin{abstract}
The celebrated Leibnitz triangle has a remarkable property, namely that each of its elements equals the sum of its South-West and South-East neighbors. In probabilistic terms, this corresponds to a specific form of correlation of  
$N$ equally probable 
binary variables which satisfy {\it
 scale-invariance}. Indeed, the marginal probabilities of the $N$-system precisely coincide
with the joint probabilities of the $(N-1)$-system.  On the other hand, the nonadditive entropy
$S_q\equiv\frac{1-\int_{-\infty}^{\infty} [p(x)]^q}{q-1}$ $(q \in {\mathbb  R};\, S_1
=-\int_{-\infty}^\infty p(x) \ln p(x))$, which grounds nonextensive statistical mechanics, is,
under appropriate constraints, extremized by the ({\it $q$--Gaussian}) distribution $p_q(x)
\propto [1-(1-q) \, \beta \, x^2]^{1/(1-q)}$ ($q<3$; $p_1(x) \propto e^{-\beta   x^2})$.  These
distributions also result, as attractors, from a generalized central limit theorem for random variables which have a finite generalized variance, and are correlated in a specific way called {\it $q$--independence}. 
In order to physically enlighten this concept, we introduce here three types of asymptotically scale-invariant probabilistic models with binary random variables, namely (i) a family, characterized by an index $\nu=1,2,3,\dots$, unifying  the Leibnitz triangle ($\nu=1$) and the case of independent variables ($\nu\to\infty$); (ii) two slightly different discretizations of $q$--Gaussians; (iii) a special family, characterized by the parameter $\chi$, which generalizes the usual case of independent variables (recovered for $\chi=1/2$). Models (i) and (iii) are in fact strictly scale-invariant.   
For models (i), we analytically show that the $N \to\infty$ probability distribution is a $q$--Gaussian with $q=(\nu -2)/(\nu-1)$. Models (ii) approach $q$--Gaussians by construction, and we numerically show that they do so with asymptotic scale-invariance. Models (iii), like two other strictly scale-invariant models recently discussed by Hilhorst and Schehr (2007), approach instead limiting distributions which are {\it not} $q$--Gaussians. 
The scenario which emerges is that asymptotic (or even strict) scale-invariance is not sufficient but it might be necessary for having strict (or asymptotic) $q$--independence, which, in turn, mandates $q$--Gaussian attractors.
\end{abstract}

\maketitle

\section{INTRODUCTION}
\label{introduction}

The central limit theorem (CLT) provides  a most powerful tool to explain the ubiquity of
Gaussian distributions in physical systems. It explains that $N$ independent or weakly
correlated arbitrarily distributed random variables, with finite variances, sum up to Gaussian
probability distributions for $N \to \infty$, corresponding to the thermodynamical limit in physical systems. This theorem constitutes part of the
foundations of Boltzmann-Gibbs (BG) statistical mechanics, making it possible to describe a
vast number of systems without accounting for the specific micro-dynamics constituting them.  

On the other hand, in systems dominated by strongly correlated microscopic events, correlations have remained a stumbling block for the researcher, making it extremely difficult to adequately take into account the contribution of all the microscopic events in order to get a general macroscopic behavior. 

Recently, specific generalizations of the CLT have been proposed taking into account some
classes of global correlations, typically  correlations over long
distances~\cite{Tsallis:05,Umarov:08,Umarov:06a,Umarov:07,UmarovTsallis2007,UmarovTsallis2008,UmarovQueiros2008,TsallisQueiros2007,QueirosTsallis2007}.
Let us briefly review the present situation. When we have $N$ identical 
{\it independent} random variables whose individual distribution has a {\it finite} variance,
their sum approaches, as $N$ diverges and after appropriate centering and scaling, a {\it
  Gaussian distribution}. This is the so-called standard CLT. If the individual variance {\it
  diverges} (due to fat tails of the power-law class, excepting for possible logarithmic corrections), the attractor is a {\it L\'evy distribution} (also called sometimes {\it
  $\alpha$-stable distribution}).  If the variables are not independent but {\it
  $q$--independent} (the $q=1$ particular instance recovers standard probabilistic
independence), then, if certain $q$--generalized variance is {\it finite}, the attractor is a
{\it $q$--Gaussian distribution} \cite{Umarov:08}; if that variance {\it diverges} (due to specific power-law asymptotics), then the
attractor is a {\it $(q,\alpha)$-stable distribution} \cite{Umarov:06a}. These various results
have been numerically illustrated in \cite{TsallisQueiros2007,QueirosTsallis2007}, and some
extensions can be seen in
\cite{Umarov:07,UmarovTsallis2007,UmarovTsallis2008,UmarovQueiros2008,Vignat:07}. When $q=1$,
the correlations disappear, and the $q$--Gaussian  ($(q,\alpha)$-stable distribution)
reproduces a Gaussian (L\'evy distribution). In terms of mathematical grounding of statistical
mechanics, the $q \ne 1$ CLT cases play for nonextensive statistical mechanics
\cite{Tsallis:88,next:04,next:05,biblio} the same role that the $q=1$ CLT cases play for the BG
theory. The development of this theory is motivated by the observation that $q$-Gaussians (or distributions very close to them) 
appear in many real physical systems, such as  cold atoms in
dissipative optical lattices \cite{Douglas2006}, dusty plasma \cite{Goree2008}, motion of Hydra cells \cite{Upaddhyaya:01}, and defect turbulence \cite{Daniels:04}. This suggests that this kind of  
probability distribution plays an important role in system out-of equilibrium presenting global correlations.

A central point of this generalized theorem is of course the hypothesis of $q$--independence (defined in \cite{Umarov:08} through the $q$--product \cite{Nivanen:03,Borges:04}, and the $q$--generalized Fourier transform \cite{Umarov:08}). This corresponds, when $q \ne 1$, to a global correlation of the $N$ random variables. Its rigorous definition is however not transparent enough in physical terms. An important goal along these lines is therefore to describe in simple terms the basic physical assumptions behind the mathematical requirement of $q$--independence. Two types of simple models have been recently introduced \cite{MoyanoGellmann:2006,Thistleton:06} in order to provide this insight. They are hereafter referred to as the MTG and the TMNT models respectively. The first one is a discrete model with binary random variables; the second one involves continuous variables. They both are strictly scale-invariant and have been  found numerically to converge, when $N$ increases, on distributions remarkably close to $q$--Gaussians. However, a rigorous analytical
treatment showed that the functional form of the probability distribution, although being amazingly well fitted by $q$--Gaussians in both models, differs from it in the thermodynamical limit~\cite{Hilhorst:07}. This fact established that strict scale-invariance (hence asymptotic-scale invariance) is not sufficient for having $q$--Gaussians as limiting distributions. It remained open the question whether scale-invariance allows for such limiting distributions. In the present paper, we precisely clarify this central issue, thus providing  some insight into this problem.

After some brief review of the theoretical frame within which $q$--Gaussians emerge, we
introduce, in Sec.~\ref{sec:family}, a {\it strictly} scale-invariant family of Leibnitz--like
triangles, for which the limiting distribution can be exactly obtained. Its limiting
distributions are $q$--Gaussians, which 
proves that scale-invariance is consistent with $q$--Gaussianity. In Sec.~\ref{sec:discretized}
we discretize (in two slightly different manners) $q$--Gaussians. We then show numerically that
these discretized distributions approach the limiting ones ---$q$--Gaussians by construction---
with 
asymptotic scale-invariance. This illustrates that both strict and asymptotic scale-invariances are compatible with $q$--Gaussianity. Finally, in   
Sec.~\ref{sec:another} we introduce another family of strictly scale-invariant probabilistic triangles which, like the MTG and TMNT models, do not converge onto $q$--Gaussians, but onto rather curious distributions, having a singular behavior at infinity. We conclude in Sec.~\ref{conclusions}.

\section{$q$--GAUSSIANS}

In addition to their appearance in a $q$--generalized CLT, $q$--Gaussians are the nonextensive statistical
mechanical~\cite{Tsallis:88,next:04,next:05,biblio} analog to Gaussians in the
BG theory. 
By introducing a generalized entropic functional, a generalized thermostatistics could be
developed that exhibits a thermodynamic scenario similar to that of the original one. This theory accounts
for a class of systems where the BG theory fails. The entropy $S_q$ (with $q \ne 1$) was proposed as an alternative to the BG entropy for complex systems, e.g., trapped in nonergodic
non--equilibrium states~\cite{Pluchino:07,Pluchino:08}, or nonlinear dynamical systems at the
edge of chaos \cite{Tirnakli:07,Tirnakli:08}. Also, a connection between this generalized entropy and the
generalized nonlinear Fokker-Planck equation leading to anomalous diffusion has been established \cite{Bukman:96}.

Indeed,  $p_1(x)\propto e^{-x^2/2\sigma^2}$ optimizes the BG entropy
$S_1=-k\int_{-\infty}^\infty p_1(x)\ln p_1(x)\,dx$, under constraints $\int_{-\infty}^\infty
dx\, p_1(x)=1$ and $\langle x^2\rangle_1\equiv\int_{-\infty}^\infty
dx\,x^2p_1(x)=\sigma^2$. Analogously,  $q$--Gaussians  
\begin{equation}
p_q(x)\propto\left[1-(1-q)\beta x^2\right]^{\frac{1}{1-q}}~~~~~ (q<3)
\label{q-Gaussians}
\end{equation}
optimize the entropy
$$S_q[p_q(x)]=k\dfrac{1-\displaystyle\int_{-\infty}^\infty dx\, \left[p_q(x)\right]^q}{q-1}$$
under constraints $\int_{-\infty}^\infty dx\, p_q(x)=1$ and  \cite{Prato:99}
$$\langle x^2\rangle_q\equiv\dfrac{\displaystyle\int_{-\infty}^\infty dx\,
  x^2[p_q(x)]^q}{\displaystyle\int_{-\infty}^\infty dx\, [p_q(x)]^q}=\sigma^2$$    
It must be noted that $q$--Gaussians \eqref{q-Gaussians} have compact support ($|x|\leqslant
1/\sqrt{(1-q)\beta}$) for $q<1$ and are defined for all $x$ for $q\geqslant 1$. In addition,
the second moment of $q$--Gaussians remains finite for $q<5/3$. In the following, we will
consider $\beta=1$. 
The distribution 
$P_q(x) \equiv [p_q(x)]^q / \int_{-\infty}^\infty dx\, [p_q(x)]^q$ 
is called {\it escort distribution} \cite{escort} and its relevance is discussed, for instance,
in \cite{TsallisPlastinoAlvarez2008}. $S_q$ is {\it nonadditive} for $q\ne 1$ since, for two
independent systems $A$ and $B$, we easily verify (assuming $k=1$)
$S_q(A+B)=S_q(A)+S_q(B)+(1-q)S_q(A)S_q(B)$.

In complex cases, the BG entropy generally looses its extensivity, i.e., it no more (asymptotically) increases linearly with the
system size. In this paper, the emphasis is to consider $q$--Gaussians as being limiting
probability functions characteristic for non-equilibrium states.  
We will determine the characteristic entropic index for the simple systems described in the
following sections but do not necessarily expect them to yield extensivity of the $q$--entropy with the same value for $q$ that they exhibit in the stationary state distribution.

\section{FIRST MODEL: A FAMILY OF LEIBNITZ-LIKE TRIANGLES}
\label{sec:family}

In a probabilistic context,  {\it scale invariance} will be said to (strictly) occur when, for a set of $N$ random
variables, the functional form of the associated {\it marginal} probabilities of the $N$-variables set
coincides with the {\it joint} probabilities associated with the $(N-1)$-variables set, i.e, when
\begin{equation}
\int p_N(x_1,x_2,\dots,x_{N-1},x_N)\,dx_N=p_{N-1}(x_1,x_2,\dots,x_{N-1})~.  
\label{eq:scale_inv}
\end{equation}
This relation is always valid for independent random variables, where the joint probability
corresponds to the product of the individual probabilities, but it is by no means necessarily valid for
correlated ones (see, for instance, Sec. \ref{sec:discretized} for a counter example). 

We take now the case of a set of binary independent variables, each one taking
values 1 or 0 with probabilities $p$ and $1-p$ respectively. For $N=2$, the joint and marginal
probability distributions are given by Table~\ref{N=2}.

\begin{table}[ht!]
\begin{center}
\begin{tabular}{c|c|c|c}
$x_1\backslash ^{\displaystyle x_2}$ &1       &0       &\\\hline
1&$p^2$   &$p(1-p)$ &$p$\\\hline
0&$p(1-p)$&$(1-p)^2$&$1-p$\\\hline
 &$p$     &$1-p$    &1\\
\end{tabular}
\end{center}
\caption{Joint probability distribution for a set of $N=2$ independent binary variables.\label{N=2}} 
\end{table}

Last row (column) of Table \ref{N=2} represents the marginal probabilities of $x_2$ ($x_1$)
which reproduce the form of the probability distribution for each single ($N=1$) variable. For the
$N=3$ case, it is necessary to project a cube in the plane in order to represent the whole set
of probabilities (Table~\ref{N=3}). 

\begin{table}[h]
\begin{center}
\begin{tabular}{c|c|c|}
${x_1\backslash^{\displaystyle x_2}}^{\displaystyle\backslash^{\displaystyle x_3}}$ &1       &0       \\\hline
\multirow{2}{1cm}{1}&$p^3$        &$p^2(1-p)$ \\
                 &$[p^2(1-p)]$   &$[p(1-p)^2]$ \\\hline
\multirow{2}{1cm}{0}&$p^2(1-p)$    &$p(1-p)^2$\\                                &$[p(1-p)^2]$&$[(1-p)^3]$\\\hline
\end{tabular}
\end{center}
\caption{Joint probability distribution for a set of $N=3$ independent binary variables.\label{N=3}} 
\end{table}
Each box of Table \ref{N=3} contains two probabilities. The one in brackets stands for the case
$x_3=0$, the other being for $x_3=1$. Adding up the two probabilities of each box of Table
\ref{N=3} we get the corresponding box of Table \ref{N=2}, so scale invariance,
  eq.~(\ref{eq:scale_inv}), comes up again, 
as it does when increasing $N$. 

It is clear that among the $2^N$ elementary events of the sample space, only $N+1$ have
different probabilities $r_{N,n}=p^{N-n}(1-p)^n$, for $n=0,\dots,N$, which, as a function of
$N$, can be displayed in a triangle in the form 

\begin{tabular}{cccccccccccc}
$(N=0)$&&&&&&1&&&&\\
$(N=1)$&&&&&$p$&&$1-p$&&&&\\
$(N=2)$&&&&$p^2$&&$p(1-p)$&&$(1-p)^2$&&&\\
$(N=3)$&&&$p^3$&&$p^2(1-p)$&&$p(1-p)^2$&&$(1-p)^3$&&\\
&&&&&$\vdots$&&$\vdots$&&&&
\end{tabular}

The probabilities $r_{N,n}$ are the joint $N$-variable probabilities.
The above triangle reflects the aforementioned scale invariance,
eq.(\ref{eq:scale_inv}), in the sense that its 
coefficients satisfy the relation 
\begin{equation}\label{scale_invariance}
r_{N,n}+r_{N,n+1}=r_{N-1,n}\end{equation}
that is, the sum of two consecutive coefficients (marginal probabilities of a
  $N$--system) in the same row yields the coefficient on top
of them (joint probabilities of the $(N-1)$--system). In other words, the corresponding marginal probabilities happen to coincide with the row just
above, a quite remarkable property (by no means general: see Sec. \ref{sec:discretized}). More precisely, we are comparing two systems: one with $N$ elements and one with ($N-1$) 
elements. And eq. (\ref{scale_invariance}) means that the probabilistic observation of the ($N-1$)-system coincides with the observation of ($N-1$) particles of the $N$-system. Relation \eqref{scale_invariance} can alternatively be given as a rule to generate the
triangle together with the starting condition $r_{N,0}=p^N$ for each row.   
 
Let us focus now on the random variable $z=x_1+x_2+\cdots+x_N$. It takes the values
$0,1,\dots,N$, with a degeneracy ---imposed by the identical character of the $N$ binary
subsystems--- given by the binomial coefficients $\binom{N}{n}$, so the actual set
of probabilities for $z$  
\begin{equation}\label{probabilities}p_{N,n}\equiv P(z=N-n)=\binom{N}{n}r_{N,n}\end{equation}are to be calculated multiplying the above triangle by the Pascal triangle. It can be easily
verified that $p_{N,n}$ is the binomial distribution which has, as limiting probability function
($N\to \infty$), a Gaussian.

Scale invariance condition \eqref{scale_invariance} is the so called {\it Leibnitz triangle rule}. The Leibnitz triangle

\begin{tabular}{ccccccccccccccccc}
$(N=0)\quad\quad$&&&&&&&&1&&&&&&&\\
$(N=1)\quad\quad$&&&&&&&$\dfrac{1}{2}$&&$\dfrac{1}{2}$&&&&&&\\
$(N=2)\quad\quad$&&&&&&$\dfrac{1}{3}$&&$\dfrac{1}{6}$&&$\dfrac{1}{3}$&&&&&\\
$(N=3)\quad\quad$&&&&&$\dfrac{1}{4}$&&$\dfrac{1}{12}$&&$\dfrac{1}{12}$&&$\dfrac{1}{4}$&&&&\\
$(N=4)\quad\quad$&&&&$\dfrac{1}{5}$&&$\dfrac{1}{20}$&&$\dfrac{1}{30}$&&$\dfrac{1}{20}$&&$\dfrac{1}{5}$&&&\\
$(N=5)\quad\quad$&&&$\dfrac{1}{6}$&&$\dfrac{1}{30}$&&$\dfrac{1}{60}$&&$\dfrac{1}{60}$&&$\dfrac{1}{30}$&&$\dfrac{1}{6}$&&\\
&&&&$\vdots$&&&&$\vdots$&&&&$\vdots$&
\end{tabular}

\noindent
satisfies condition \eqref{scale_invariance} and differs from the independent case in the
definition of the starting condition, now given by $r_{N,0}=\dfrac{1}{N+1}$. Leibnitz triangle
coefficients may be interpreted as a way to introduce correlations in the $N$ random variables
system. Probabilities are again calculated with \eqref{probabilities}, 
i. e., by multiplying Leibnitz and Pascal triangles to get
$p_{N,n}=\dfrac{1}{N+1}$.  Hence, the Leibnitz triangle rule leads to a
uniform probability distribution, and so can be related to a
$q$--Gaussian in the limit case $q\to-\infty$.

We will now generalize the Leibnitz triangle introducing a family of scale invariant triangles
$r^{(\nu)}_{N,n}$, $\nu=1,2,\dots$, with boundary coefficients given by 
\begin{eqnarray}
r^{(1)}_{N,0}&=&\frac{1}{N+1},\nonumber\\
r^{(2)}_{N,0}&=&\frac{2\cdot 3}{(N+2)(N+3)}~,\nonumber\\ 
r^{(3)}_{N,0}&=&\frac{3\cdot 4\cdot 5}{(N+3)(N+4)(N+5)}~,~~...~~,~ \nonumber\\ r^{(\nu)}_{N,0}&=&\frac{\nu\cdot\cdot\cdot(2\nu-1)}{(N+\nu)\cdot\cdot\cdot(N+2\nu-1)}~.
  \label{eq:F_postul}
\end{eqnarray}
which recovers the Leibnitz triangle for $\nu=1$. Let us emphasize that this definition leads to i) positive, ii) symmetric, and iii) norm preserving (in the sense that $\sum_{n=0}^Np_{N,n}=1$, for all values of $N$) triangles. As a second example, the triangle for $\nu=2$ reads

\begin{tabular}{ccccccccccccccccc}
$(N=0)\quad\quad$&&&&&&&&1&&&&&&&\\
$(N=1)\quad\quad$&&&&&&&$\dfrac{1}{2}$&&$\dfrac{1}{2}$&&&&&&\\
$(N=2)\quad\quad$&&&&&&$\dfrac{3}{10}$&&$\dfrac{1}{5}$&&$\dfrac{3}{10}$&&&&&\\
$(N=3)\quad\quad$&&&&&$\dfrac{1}{5}$&&$\dfrac{1}{10}$&&$\dfrac{1}{10}$&&$\dfrac{1}{5}$&&&&\\
$(N=4)\quad\quad$&&&&$\dfrac{1}{7}$&&$\dfrac{2}{35}$&&$\dfrac{3}{70}$&&$\dfrac{2}{35}$&&$\dfrac{1}{7}$&&&\\
$(N=5)\quad\quad$&&&$\dfrac{3}{28}$&&$\dfrac{1}{28}$&&$\dfrac{3}{140}$&&$\dfrac{3}{140}$&&$\dfrac{1}{28}$&&$\dfrac{3}{28}$&&\\
&&&&$\vdots$&&&&$\vdots$&&&&$\vdots$&
\end{tabular}

It may be shown that the coefficients of consecutive triangles of the family are related to each other in the way
\begin{equation}
  \label{eq:F_transrel}
  r^{(\nu)}_{N,n} = \frac{r^{(\nu-1)}_{N+2,n+1}}{r^{(\nu-1)}_{2,1}}~.
\end{equation}
Therefore, all of them can be expressed in terms of the Leibnitz triangle and so a general expression for the coefficients may be obtained
\begin{equation}
  \label{eq:F_transrel2}
  r^{(\nu)}_{N,n}=\frac{r_{N+2(\nu-1),n+\nu-1}^{(1)}}{r_{2(\nu-1),\nu-1}^{(1)}}= \frac{(2\nu-1)!}{[(\nu-1)!]^2(N+2\nu-1)\binom{N+2(\nu-1)}{n+\nu-1}}.
\end{equation}
In particular, the central elements of the triangle ($n=N/2$ for even $N$)
\begin{equation}
  \label{eq:central_elements}
  r^{(\nu)}_{N,\frac{N}{2}}= \frac{(2\nu-1)!}{[(\nu-1)!]^2(N+2\nu-1)\binom{N+2(\nu-1)}{\frac{N}{2}+\nu-1}}
\end{equation}
can be used to generate the whole triangle starting from the center instead of the side. 

We will now show that not only the Leibnitz triangle but also the rest of the triangles of the
family yield $q-$Gaussians as limiting probability distribution. In fact, there is a value
$q=q_\text{lim}(\nu)$, for which the $q_\text{lim}-$Gaussian corresponds
to the $N \to \infty$ probability distribution defined by the corresponding triangle, that is
\begin{equation}p^{(\nu)}_{N,n}=\binom{N}{n}r^{(\nu)}_{N,n}\to
  \mathcal{P}_{q_\text{lim}}(x).\label{Ptop_q}\end{equation}   
for $N\to\infty$ (as $n=0,1,...,N$, we need to define $x$ in terms of $n$ and $N$, normally
  through appropriate centering and scaling). For this purpose we will express the boundary coefficients 
$r^{(\nu)}_{N,0}$ in an alternative way by using partial fraction decomposition
\begin{equation}
  \label{eq:F_coeff_gen}
  r^{(\nu)}_{N,0} = \frac{(2\nu-1)!}{[(\nu-1)!]^2}\sum \limits_{j=0}^{\nu-1} (-1)^{j}\binom{\nu-1}{j} \frac{1}{N+\nu+j}
\end{equation}
On the other hand, due to the scale invariance rule, any term of the triangle can be expressed
as a function of the boundary terms in the form  
\begin{equation}
  \label{eq:F_coeff}
  r^{(\nu)}_{N,n} = \sum \limits^n_{i=0} (-1)^{n-i} \binom{n}{i} r^{(\nu)}_{N-i,0}~,
\end{equation}
Introducing Eq.~\eqref{eq:F_coeff_gen} in Eq.~\eqref{eq:F_coeff} yields 
\begin{equation}
\label{eq:F_coeff_gen_bis}
  r^{(\nu)}_{N,n} = \sum \limits_{j=0}^{\nu-1}a_j^{(\nu)}\sum \limits^n_{i=0} (-1)^{n-i} \binom{n}{i}   \frac{1}{N+\nu-i+j}~.
\end{equation}
where we have made the substitution $a_j^{(\nu)}\equiv\frac{(2\nu-1)!}{[(\nu-1)!]^2}(-1)^{j}\binom{\nu-1}{j}$. 

By means of the relation $\frac{1}{N+\alpha} = \int \limits_0^\infty e^{-(N+\alpha)z} dz$, together with the binomial expansion $ (a+b)^n = \sum \limits_{i=0}^n \binom{n}{i} a^{n-i} b^i$, Eq.~\eqref{eq:F_coeff_gen_bis} can be cast in the form
\begin{align}
\label{eq:F_coeff_gen_tris}
  r^{(\nu)}_{N,n} &= \sum \limits_{j=0}^{\nu-1}a_j^{(\nu)}\int \limits_0^\infty dz\sum
  \limits^n_{i=0} (-1)^{n-i} \binom{n}{i}   e^{-(N+\nu-i+j)z}~\nonumber\\ 
  &=\sum \limits_{j=0}^{\nu-1}a_j^{(\nu)}\int \limits_0^\infty dz  e^{-(N+\nu+j)z}\sum
  \limits^n_{i=0} (-1)^{n-i} \binom{n}{i}   e^{iz}=\sum \limits_{j=0}^{\nu-1}a_j^{(\nu)}\int
  \limits_0^\infty dz  e^{-(N+\nu+j)z}(e^z-1)^n\nonumber\\ 
  &=\sum \limits_{j=0}^{\nu-1}a_j^{(\nu)}\int \limits_0^\infty dz
  e^{-(N+\nu+j)z}e^{n\ln(e^z-1)}=\sum \limits_{j=0}^{\nu-1}a_j^{(\nu)}\int \limits_0^\infty dz
  e^{-(\nu+j)z}e^{-Nf(z)}\end{align} 
where $f(z)=z-y\ln(e^z-1)$ with $y=n/N$. 

For large $N$, the integral in Eq.~\eqref{eq:F_coeff_gen_tris} can be evaluated by using the saddle point method. The minimum of $f(z)$ is
located at $z^\star=-\ln(1-y)$, with $f(z^\star)=-(1-y)\ln(1-y)-y\ln y$ and $f^{\prime\prime}(z^\star)=(1-y)/y$. Therefore, in the limit $N\to\infty$
\begin{align}
\label{eq:F_saddle}
r^{(\nu)}_{N,n} &\approx\sum
\limits_{j=0}^{\nu-1}a_j^{(\nu)}e^{-(\nu+j)z^\star-Nf(z^\star)}\int\limits_0^\infty dz
e^{-\frac{N}{2} (z-z^\star)^2 f^{\prime\prime}(z^\star)} \nonumber\\  
&\approx\sum \limits_{j=0}^{\nu-1}a_j^{(\nu)}\sqrt{\frac{2\pi y}{N (1-y)}} e^{\left(\nu+j+N(1-y)\right)
  \ln(1-y)+Ny\ln y}\nonumber\\ 
&=\sqrt{\frac{2\pi}{N}}\sum \limits_{j=0}^{\nu-1}a_j^{(\nu)} (1-y)^{\nu+j+N(1-y)-\frac{1}{2}}
y^{Ny+\frac{1}{2}}~. 
\end{align}
Concerning the limiting probability distribution, it can be obtained from the triangle coefficients through
\begin{align}
  \label{eq:F_limit}
  \mathcal{P}^{(\nu)}(y) = Np^{(\nu)}_{N,n}=N \binom{N}{n} r^{(\nu)}_{N,n} \approx \sqrt{\frac{N}{2\pi}}
  (1-y)^{-N(1-y)-\frac{1}{2}}y^{-(Ny+\frac{1}{2})}r^{(\nu)}_{N,n} \,,
\end{align}
where we have made use of the Stirling approximation. Inserting now 
Eq.~\eqref{eq:F_saddle} in Eq.~\eqref{eq:F_limit} yields
\begin{equation}
  \label{eq:F_fin_c}
  \mathcal{P}^{(\nu)}(y) \approx\sum \limits_{j=0}^{\nu-1}a_j^{(\nu)}(1-y)^{\nu+j-1}
\end{equation}
The largest exponent of $y$ in the distribution \eqref{eq:F_fin_c} is $2(\nu-1)$. Hence,
comparing with Eq.~\eqref{q-Gaussians}, the value of $q_\text{lim}$ for the $q$--Gaussian
limiting distribution function can be obtained doing $1/(1-q_\text{lim}) = \nu-1$,
i.e. \begin{equation}q_\text{lim}=\frac{\nu-2}{\nu-1},\label{qnu}\end{equation}  
which implies a width of the compact support given by $\Delta \equiv 2 \sqrt{1/(1-q_\text{lim})}=2\sqrt{\nu-1}$. Equation~\eqref{qnu} can be re-written as
$q_\text{lim}= 1 -1/(\nu-1)$, which reminds many analogous relations existing in the literature, such as $q_\text{entropy}=1 - 1/d$ in~\cite{Tsallis:05,Tsekouras:04,Anteneodo:04}.

The variable $y$ is defined between 0 and 1. 
However, we are interested in a centered distribution function defined within
$[-\Delta/2,\Delta/2]$ and thus apply the transformation
$x=2\sqrt{\nu-1}(y-1/2)$. The limiting function now results
\begin{align}
  \label{eq:F_limit2}
  \mathcal{P}^{(\nu)}(x)&= \dfrac{1}{2\sqrt{\nu-1}}\mathcal{P}^{(\nu)}(y) \approx
  \dfrac{1}{\sqrt{\nu-1}}\sum \limits_{j=0}^{\nu-1}\frac{a_j^{(\nu)}}{2^{j+\nu}}\left(1-\frac{x}{\sqrt{\nu-1}}\right)^{\nu+j-1}\nonumber\\
  &=\frac{(2\nu-1)!}{2^\nu\sqrt{\nu-1}[(\nu-1)!]^2} \left(1-\frac{x}{\sqrt{1-\nu}}\right)^{\nu-1}
\sum \limits_{j=0}^{\nu-1} \binom{\nu-1}{j}\frac{(-1)^{j}}{2^{j}}
\left(1-\frac{x}{\sqrt{\nu-1}}\right)^{j}\nonumber\\ 
& =\frac{(2\nu-1)!}{2^{2\nu-1}\sqrt{\nu-1}[(\nu-1)!]^2} \left(1-\frac{x^2}{\nu-1}\right)^{\nu-1},\end{align}
which exactly coincides with a $q-$Gaussian with $q=q_\text{lim}$. 

In addition, $\mathcal{P}^{(\nu)}(x)$ transforms into the Gaussian distribution for $\nu
\to\infty$, so we recover the statistical independence case. In fact, it can be verified that taking limit in
Eq.~\eqref{eq:F_transrel2} one gets $\lim_{\nu\to\infty}r^{(\nu)}_{N,n}=2^{-N}$, hence $\lim_{\nu\to\infty}p^{(\nu)}_{N,n}=\binom{N}{n} 2^{-N}$, so the
corresponding triangle is the one given at the beginning of this section with $p=1/2$. 

\begin{figure}
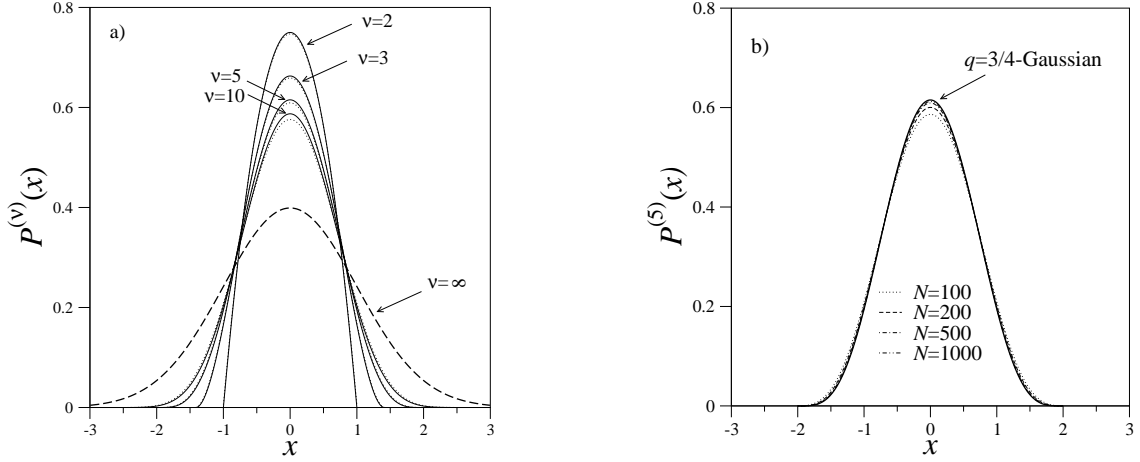

\begin{minipage}{.5\linewidth}
\epsfig{file=fig_1a.eps, width=6.5cm, clip=}
\end{minipage}\hspace*{2mm}
\begin{minipage}{.5\linewidth}
\epsfig{file=fig_1b.eps, width=6.5cm, clip=}
\end{minipage}
\caption{a) Dotted lines: Probability distributions ${\mathcal{P}^{(\nu)}(x)}=\frac{N}{2\sqrt{\nu-1}}p^{(\nu)}_{N,n}$, with $p^{(\nu)}_{N,n}$ given in \eqref{Ptop_q} for $N=500$ and $\nu=2, 3, 5$ and 10. Solid lines: Corresponding $q_\text{lim}$-Gaussians with $q_\text{lim}$ given in \eqref{qnu} for the respective values $q_\text{lim}=0, 1/2, 3/4$ and $8/9$. The limiting case $\nu\to\infty$ (Gaussian) is also depicted for comparison. b)
${\mathcal{P}^{(5)}(x)}$ for $N=100, 200, 500$ and $1000$ and the corresponding
$q_\text{lim}-$Gaussian with $q_\text{lim}=3/4$. \label{P_attractor}}
\end{figure}

Figure~\ref{P_attractor}(a) shows $\mathcal{P}^{(\nu)}(x)=\frac{N}{2\sqrt{\nu-1}}p^{(\nu)}_{N,n}$
as compared to the corresponding $q_\text{lim}-$Gaussians for $N=500$. It is apparent that the
approximation becomes poorer when increasing $\nu$. Figure~\ref{P_attractor}(b) shows the validity
of Eq.~\eqref{Ptop_q} for $\nu=5$, the corresponding  $q_\text{lim}=3/4$ and $N=$100, 200, 500
and 1000. Curves overlap with the $q-$Gaussian for greater values of $N$. The convergence is
thus evident.       

\begin{figure}[h]
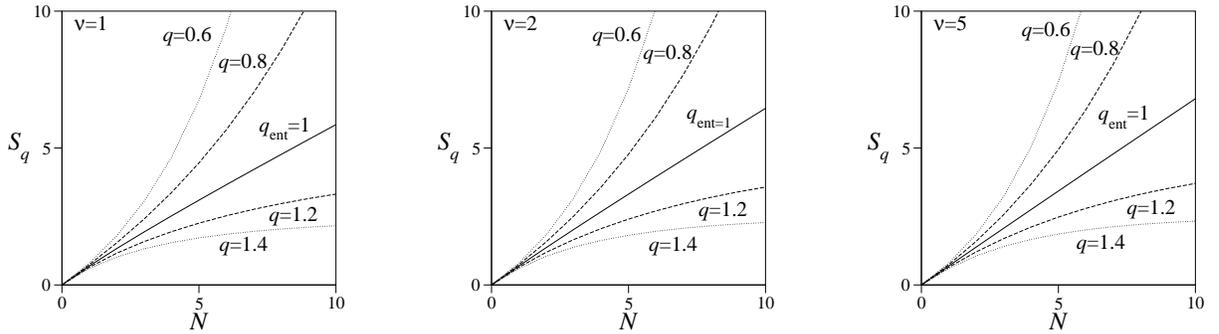

\begin{minipage}{.33\linewidth}
\epsfig{file=fig_2a.eps, width=4.5cm, clip=}
\end{minipage}\hspace*{2mm}
\begin{minipage}{.33\linewidth}
\epsfig{file=fig_2b.eps, width=4.5cm, clip=}
\end{minipage}\hspace*{2mm}
\begin{minipage}{.33\linewidth}
\epsfig{file=fig_2c.eps, width=4.5cm, clip=}
\end{minipage}
\caption{The $q-$entropy $S_q=\frac{1-\sum_{n=0}^N\binom{N}{n}\left(r_{N,n}^{(\nu)}\right)^q}{q-1}$as a function of $N$ for $\nu=1$ (left), $\nu=2$ (center), and $\nu=5$
  (right), and typical values of $q$. In all cases $q_\text{ent}=1$.\label{S_q}}
\end{figure}

Figure~\ref{S_q} shows that, in what concerns extensivity, the family of triangles \eqref{eq:F_postul} follows the
Boltzmann-Gibbs prescription, that is, the value of $q$ that makes the $q-$entropy extensive is
$q_\text{ent}=1$ for all values of $\nu$. 

As a last remark, let us associate with the $N$ random variables the variables $\sigma_i \equiv 2x_i-1= \pm 1$ ($i=1,2,...,N$), so that $\langle \sigma_i \rangle =0,\, \forall i$. We can straightforwardly prove that $\langle \sigma_i \sigma_j \rangle=\frac{1}{2\nu +1}$, $\forall i \ne j$, $\forall N$. In the limit $\nu \to \infty$ we recover $\langle \sigma_i \sigma_j \rangle=0$, as expected for independent variables, where no correlation exists.

\vspace*{\baselineskip}

\section{SECOND MODEL: DISCRETIZED $q$--GAUSSIANS}
\label{sec:discretized}

We will now introduce another probabilistic model in which we impose a priori the condition
that the $N\to\infty$ limit for the probability distribution is a $q-$Gaussian, with the aim to
study whether (strict or asymptotic) scale invariance is also obtained. In order to
  verify if the concept of $q$-independence, i.e. correlations leading to
  $q$-Gaussians, can be related to scale invariance in 
  probabilistic terms, relation \eqref{scale_invariance} is expected to be satisfied at least in the
limit $N\to \infty$, i.e., asymptotically.
 
Considering again the set of $N$ equally probable 
binary
variables, the correlations will now be given in the form  
\begin{equation}
r_{N,n}=\frac{p_q(x_{N,n})}{\displaystyle\binom{N}{n}\sum_{n=0}^Np_q(x_{N,n})}
\label{r_q-Gaussian}
\end{equation}
where $x_{N,n}$ are $N+1$ equally spaced points in the support of the $q$--Gaussian $p_q(x)$, to
be specified later. For the set of probabilities we again write $p_{N,n}=\binom{N}{n}r_{N,n}$,
which provides us with a discrete probability distribution which, by construction, follows the
shape of the   
$q$--Gaussian $p_q(x)$. 

Concerning the way to choose the points $x_{N,n}$ in
\eqref{r_q-Gaussian} a distinction must be made between cases $q<1$ and $q\geqslant 1$. 

As mentioned before, for $q<1$, $q$--Gaussians have compact and symmetric support of width $\Delta\equiv 2/\sqrt{(1-q)}$. For this case, we will consider two different ways to choose the points $x_{N,n}$ in the support of the $q$--Gaussian:

\paragraph*{1) $N+2$ discretization (D1): }
In this implementation, we take $x_{N,n}=x_\text{min}+(n+1)h$, for $n=0,1,\dots,N$, with $x_\text{min}=-\Delta/2$ and $h=\Delta/(N+2)$, i.e., explicitly, $x_{N,n} = x_{\text{min}} + h, x_{\text{min}} + 2h, x_{\text{min}} + 3h, ...,  x_{\text{min}} + (N+1)h$.

\paragraph*{2) $N+1$ discretization (D2): }
Now, the points $x_{N,n}$ are chosen differently: the same initial interval $\Delta$ breaks now into $N+1$ equal subintervals (not $N+2$ as before) of width $h=\Delta/(N+1)$ and we take the values of $p_q(x_{N,n})$ in the center of each subinterval, i.e. $x_{N,n} = x_{\text{min}} +(2n+1)h/2$. The whole set reads $x_{N,n}=x_{\text{min}}+h/2, x_{\text{min}}+3h/2, x_{\text{min}}+5h/2,..., x_{\text{min}}+(2N+1)h/2$.

In contrast, for $q\geqslant 1$, the support
for $q$--Gaussians is the whole real axis and we must take this into account in the fit. We will
take an increasing width for the fit interval in the form $\Delta_N=\delta(N+1)^\gamma$,
$\delta$ being some initial width, and $\gamma$, with $0\leqslant\gamma\leqslant 1$, a
parameter determining the growth of the interval width (for $\gamma=0$ we recover the $q<1$
case). Now $x_{N,n}=x_{\text{min},N}+(n+1)h_N$, with $x_{\text{min},N}=-\Delta_N/2$ and
$h_N=\Delta_N/(N+2)$. 

Despite the fact that different discretizations yield different triangles \eqref{r_q-Gaussian} for a given value of $q$, let us emphasize that the corresponding limiting distributions $p_{N,n}$ tend to the {\it same} $q-$Gaussian $p_q(x)$ in the limit $N \to \infty$.

Now the following question arises. Do the triangles \eqref{r_q-Gaussian} satisfy relation \eqref{scale_invariance} as the
triangles \eqref{eq:F_postul} from Sec. \ref{sec:family} do? In other words, can $q$--Gaussians be
related to strictly scale-invariant distributions? Strictly speaking, they are not, since relation \eqref{scale_invariance} is not exactly fulfilled
(except for the case $q=0$ with the first discretization D1, as we will show later), but we
will show (analytically in some cases, numerically in others) that these triangles are
{\it asymptotically} scale-invariant, that is, relation \eqref{scale_invariance} is satisfied for
$N\to\infty$, or, alternatively, the ratio 
\begin{equation}
Q_{N,n}\equiv\frac{r_{N,n}}{r_{N+1,n}+r_{N+1,n+1}}
\label{quotient}
\end{equation}  
tends to 1 (or equivalently $Q_{N,n}-1$ tends to 0) as $N$ increases. Note that $Q_{0,0}=Q_{1,0}=Q_{1,1}=1$ for arbitrary values
of $r_{0,0}$,  $r_{1,n}$ and $r_{2,n}$.

\subsection{The $q<1$ case}

\begin{figure}
\centering\epsfig{file=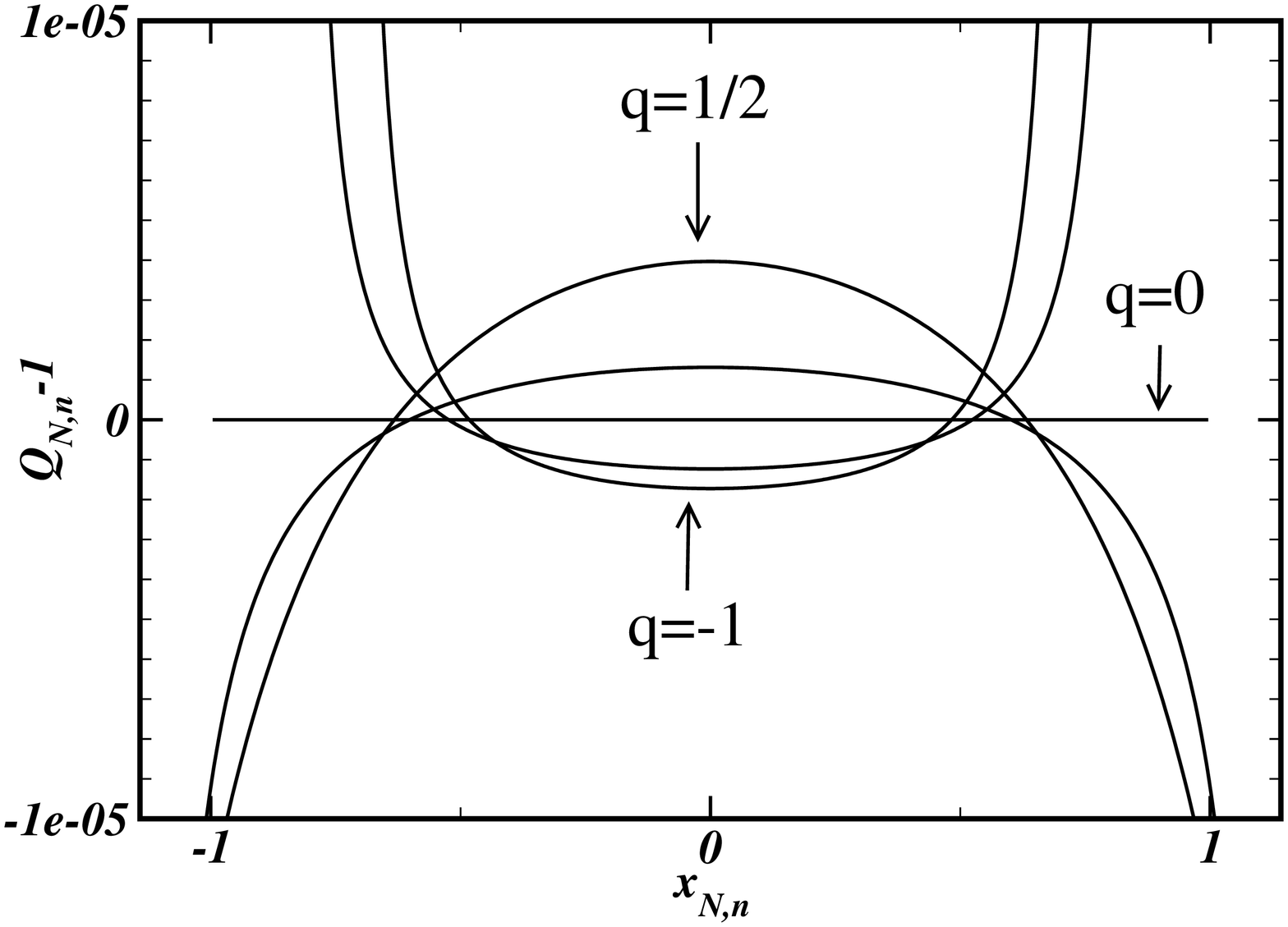, width=0.6\textwidth,angle=-90, clip=}
\centering\epsfig{file=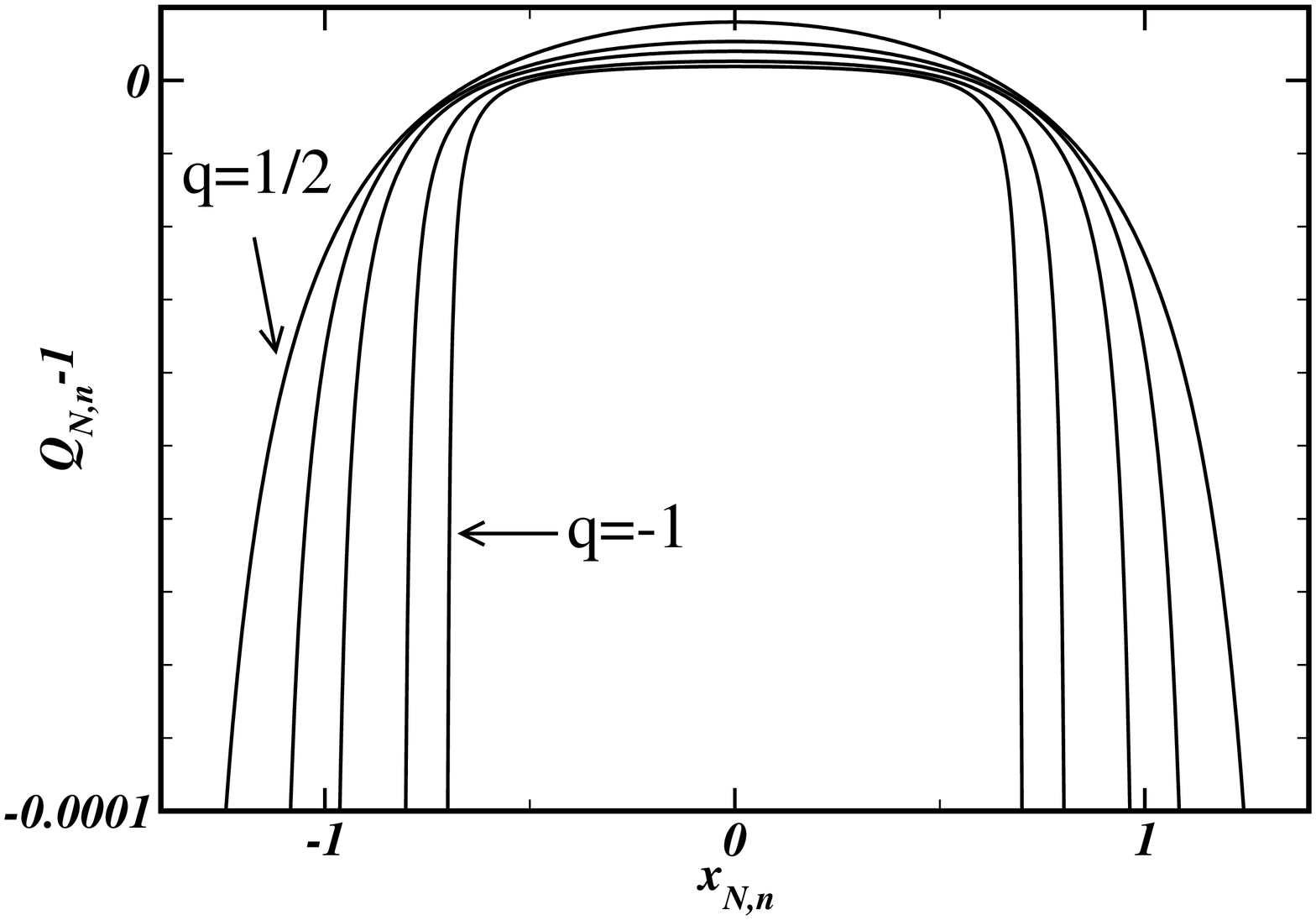, width=0.6\textwidth, angle=-90, clip=}
\caption{$Q_{N,n}-1$ as a function of $x_{N,n}$ for $N=500$ and different values of $q=-1,-1/2,0,1/4,1/2$ for discretizations D1 (top) and D2 (bottom). Strict scale invariance is observed for $q=0$ and discretization D1. In the rest of the cases deviation from zero is small and, as can be seen by results not shown here, decreases when increasing $N$.\label{figure_1}}
\end{figure}

Figure~\ref{figure_1} shows $Q_{N,n}-1$ as a function of $n$ for  $N=500$ and
different values of $q$, for both the D1 and D2--discretizations. It
is clearly observed the proximity of $Q_{N,n}$ to 1, which is more 
noticeable in the center of the triangle.

Quite remarkably, for $q=0$, strict scale invariance
is obtained in the first discretization, that is, $Q_{N,n}=1$ for all $N$ and $n$. This is so because, in this case, it can be proved that triangle \eqref{r_q-Gaussian} exactly coincides with the Leibnitz-like triangle of the family \eqref{eq:F_postul} with $\nu=2$, with associated probabilities  $p^{(2)}_{N,n}=\binom{N}{n}r^{(2)}_{N,n}=6[1+n(N-n)/(N+1)]/(N+2)(N+3)$. 

An exact expression for $r_{N,n}$ and hence $Q_{N,n}$ can also be  obtained for the D2 discretization and $q=0$, the probabilities being in this case $p_{N,n} =
3[(2n+1)(2N-2n+1)]/[(N+1)(2N^2+4N+3)]$. Of particular interest are the central
value, $Q_c\equiv Q_{N,N/2}$, and the boundary one, $Q_0\equiv Q_{N,0}$, of quotient \eqref{quotient}, being given by
\begin{eqnarray}
  \label{eq:quotients}
  Q_c & = & \frac{(N+2)^2N(2N^2+8N+9)}{(2N^2+4N+3)(N^3+6N^2+10N+2)}~~~\qquad \text{for odd $N$,}\\
\label{eq:quotients2}
  Q_0 & = & \frac{(2N+1)(N+2)(2N^2+8N+9)}{(2N^2+4N+3)(2N^2+11N+6)}~.
\end{eqnarray}

\begin{figure}
\centering\epsfig{file=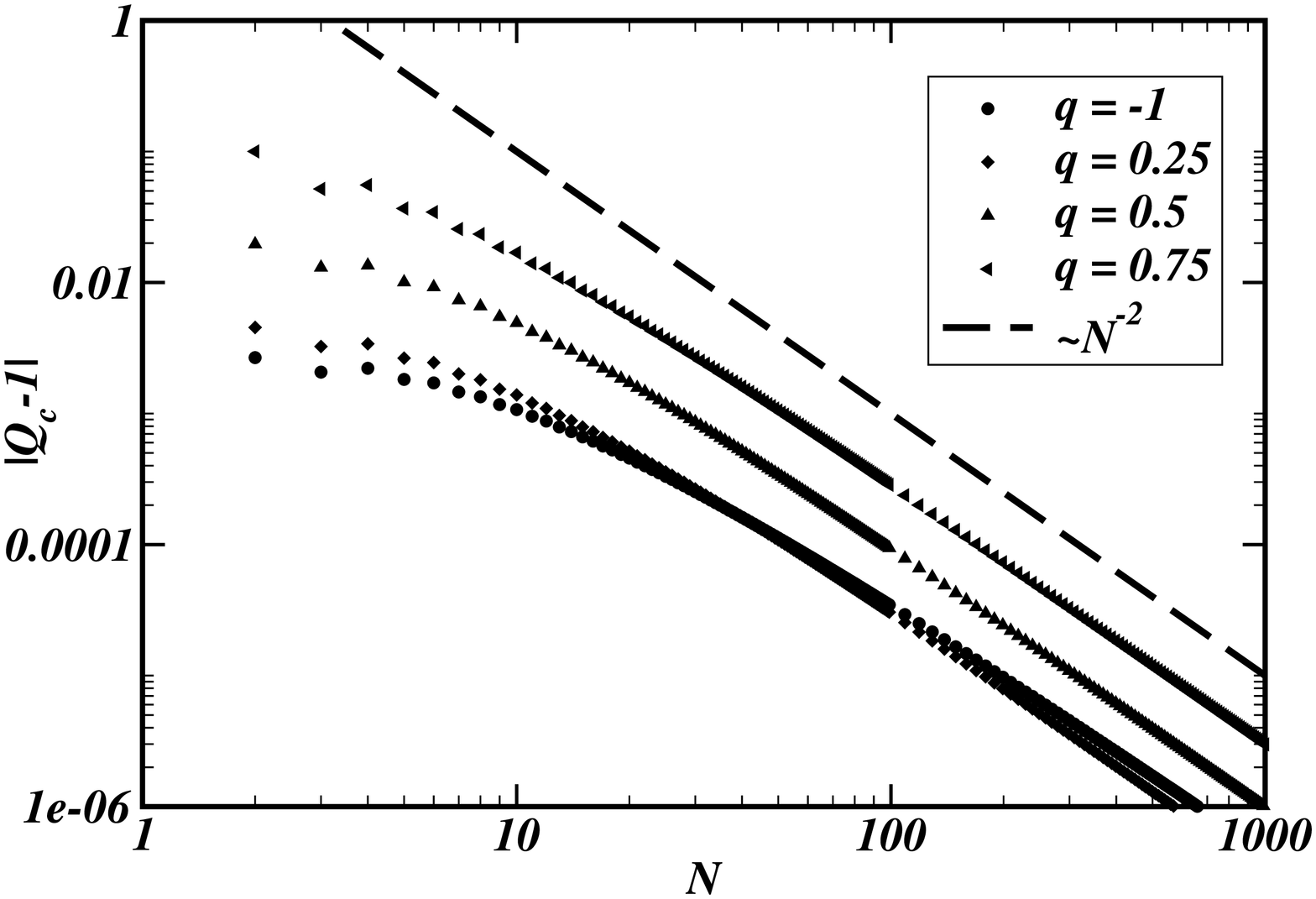, width=0.6\textwidth, angle=-90, clip=}
\centering\epsfig{file=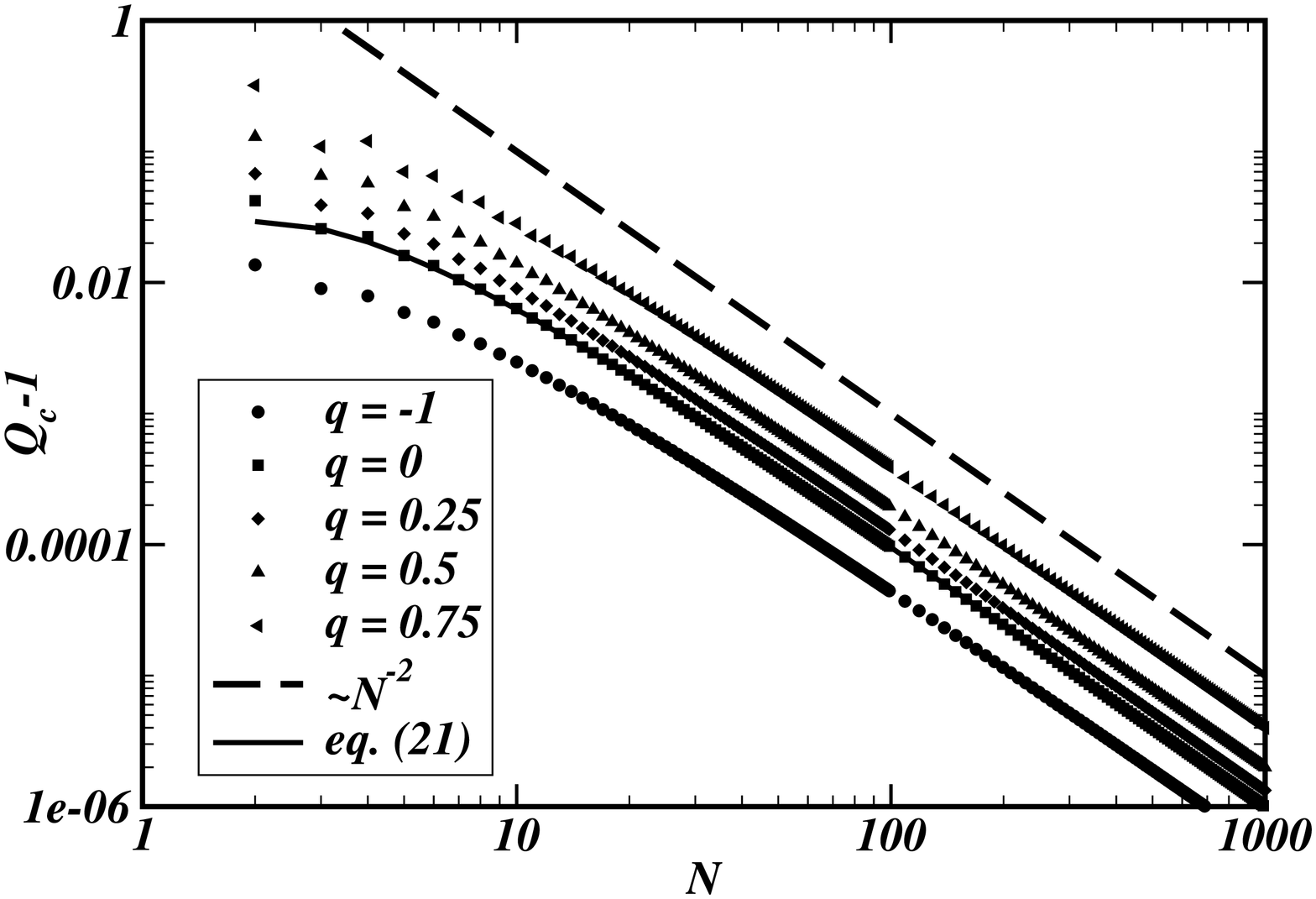, width=0.6\textwidth, angle=-90, clip=}
\caption{Central ratio $Q_{c}-1$ as a function of $N$ for discretized $q-$Gaussians with $q=-1, 0, 1/4, 1/2$ and $3/4$ for discretizations D1 (top) and D2 (bottom). Exact result \eqref{eq:quotients} for $q=0$ is also shown for D2 discretization. The power law with exponent $-2$ is shown for comparison. 
}
\label{figure_2}
\end{figure}

From Eqs.~\eqref{eq:quotients} and \eqref{eq:quotients2} results  $Q_c-1\sim N^{-2}$ and
$Q_0-1\sim -N^{-1}$, respectively. Though this equations are only valid for $q=0$, this trend
is observed for any value of $q<1$.  
Figure~\ref{figure_2} shows in a log-log plot $Q_c-1$ as a function of $N$ for
different values of $q<1$ and both discretizations. It is clear that
the decay follows a $1/N^2$ power-law for large $N$ and any value of $q$. No
substantial differences are observed between both discretizations.

\begin{figure}
\centering\epsfig{file=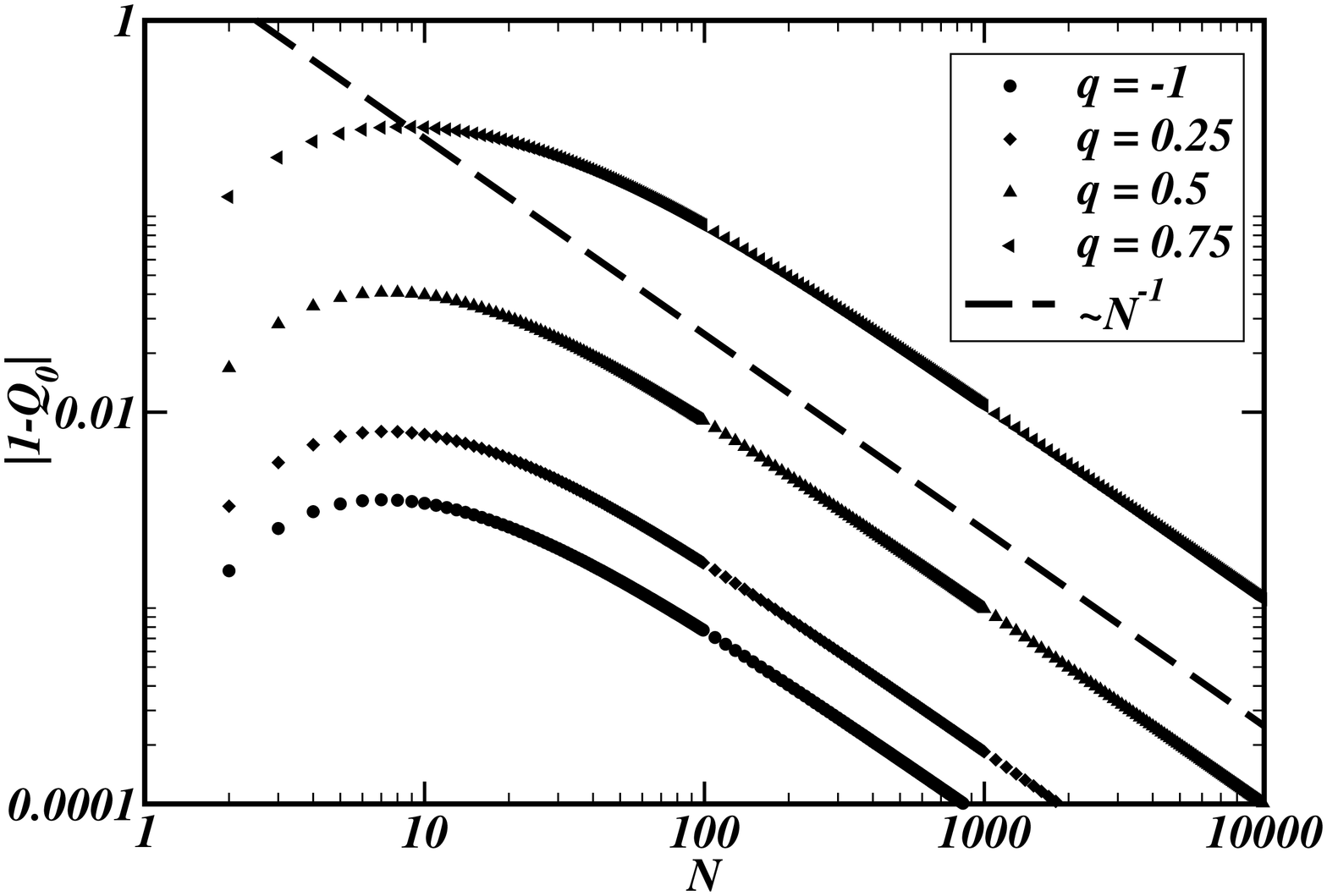, width=0.6\textwidth, angle=-90, clip=}
\centering\epsfig{file=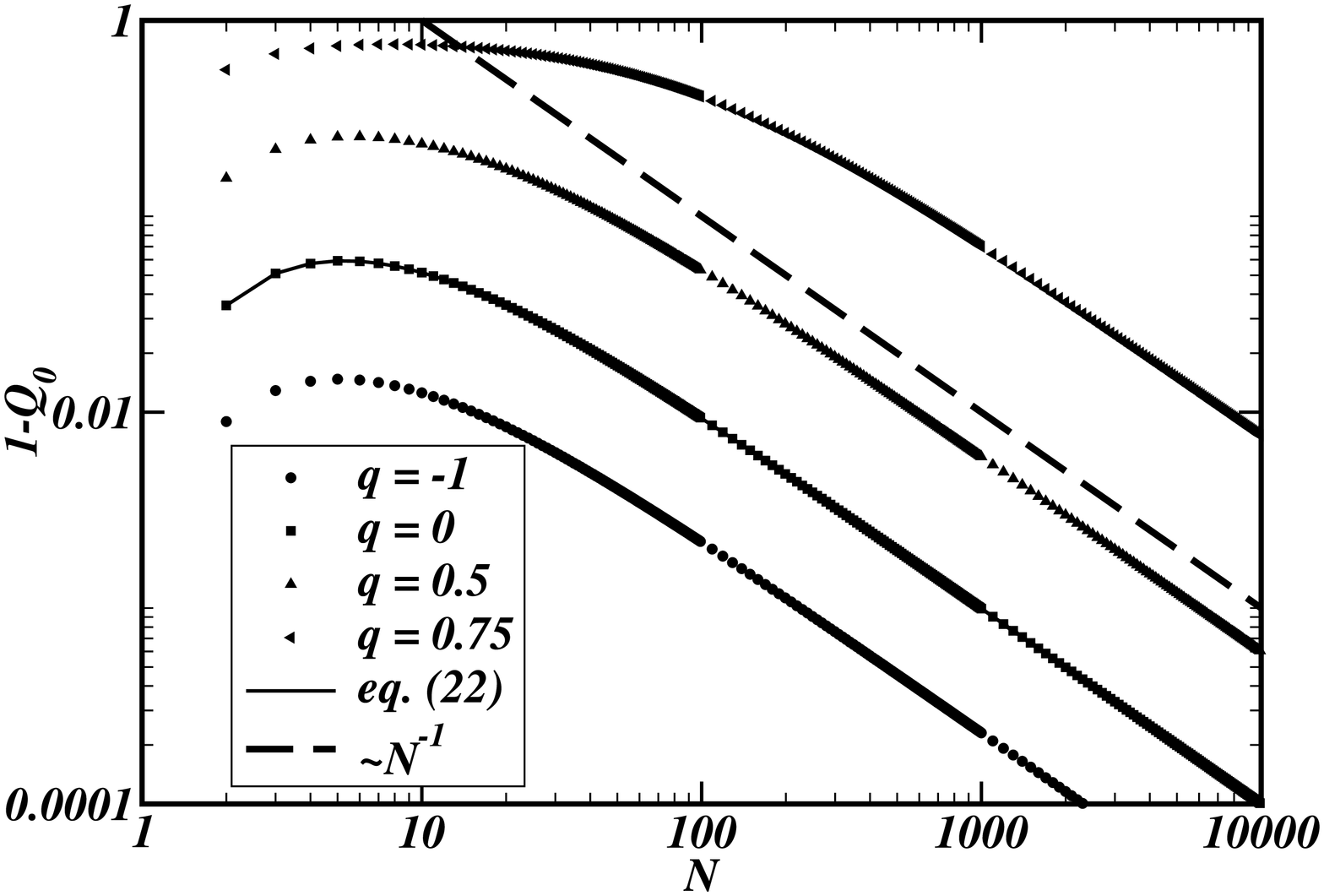, width=0.6\textwidth, angle=-90, clip=}
\caption{Boundary ratio $Q_0-1$ as a function of $N$ for discretized $q-$Gaussians with $q=-1, 0, 1/4, 1/2$ and $3/4$ for discretizations D1 (top) and D2 (bottom). Exact result \eqref{eq:quotients2} for $q=0$ is also shown for D2 discretization. The power law with exponent $-1$ is shown for comparison.
\label{figure_3}}
\end{figure}

Analogously, Fig.~\ref{figure_3} shows $Q_0-1$ as a function of $N$ and different values of $q$. We observe now a $1/N$ power-law. We found that this $1/N$ power-law transforms into the $1/N^2$ when we do not take into account the complete interval of the compact support of the $q$--Gaussian under consideration (not shown).

\begin{figure}[h]
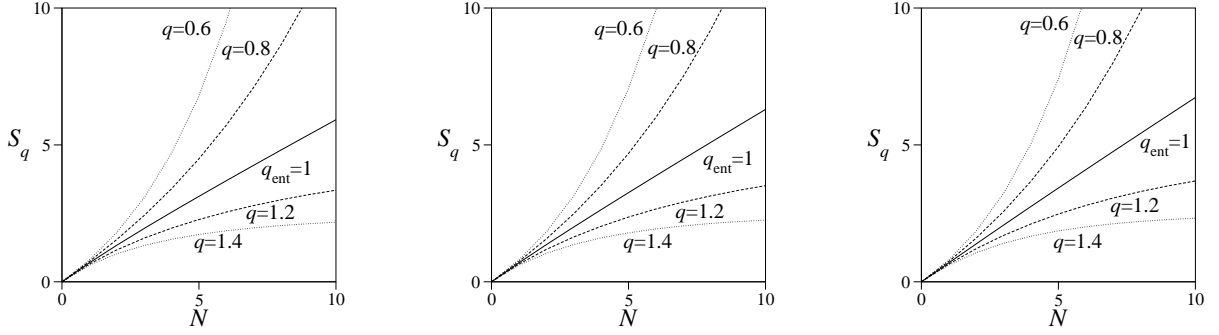

\begin{minipage}{.33\linewidth}
\epsfig{file=fig_6a.eps, width=4.5cm, clip=}
\end{minipage}\hspace*{2mm}
\begin{minipage}{.33\linewidth}
\epsfig{file=fig_6b.eps, width=4.5cm, clip=}
\end{minipage}\hspace*{2mm}
\begin{minipage}{.33\linewidth}
\epsfig{file=fig_6c.eps, width=4.5cm, clip=}
\end{minipage}
\caption{ $q-$entropy $S_q=\frac{1-\sum_{n=0}^N\binom{N}{n}r_{N,n}^q}{q-1}$, with $r_{N,n}$ given in \eqref{r_q-Gaussian}, as a function of $N$ for discretized $q$--Gaussians with $q=-1$ (left), $q=-1/2$
  (center), and $q=1/2$ (right) and discretization D1. Results with discretization D2 are indistinguishable. In all cases $q_\text{ent}=1$. \label{S_q_q-G}}
\end{figure}

Concerning the extensivity of $S_q$, the same behavior as in the previous systems is found. We get $q_\text{ent}=1$ no matter the value of $q\leqslant 1$ of the discretized $q$--Gaussian (let us remind that there is no reason for the value of $q$ of the discretized $q-$Gaussian be equal, or even simply related, to the index $q_\text{ent}$ corresponding to the extensivity of the entropy $S_q(N)$), and no matter the type of discretization (D1 or D2). Figure~\ref{S_q_q-G} shows the $q-$entropy for discretized $q$--Gaussians for typical values of $q$. The results are independent of the discretization.

\subsection{The $q\geqslant 1$ case}

\begin{figure}[h!]
\centering\epsfig{file=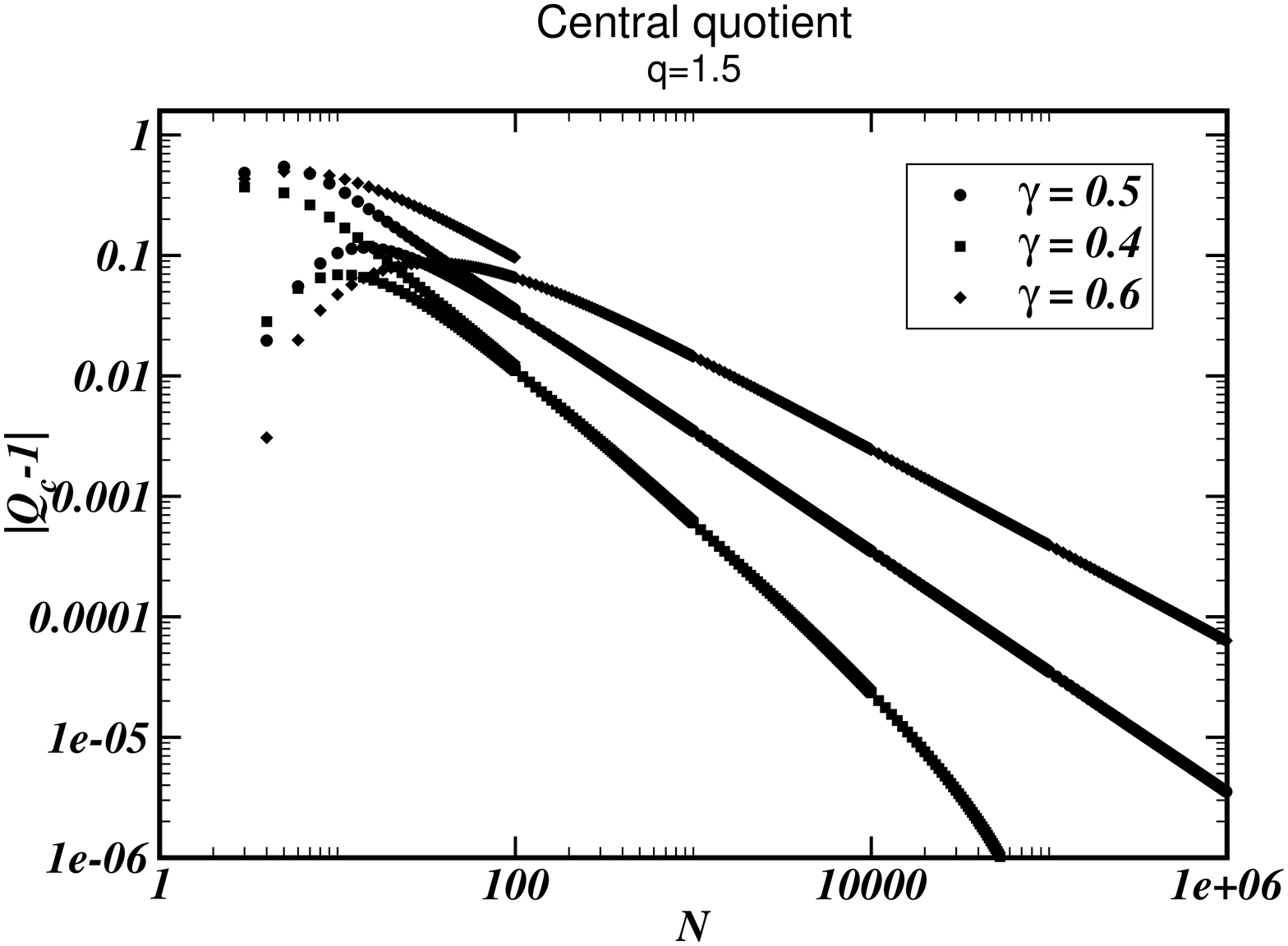,
width=0.6\textwidth,angle=-90,clip=}
\caption{Central ratio $|Q_c-1|$ as a function of $N$ for $q=3/2$, $\delta=2$ and $\gamma=$0.4, 0.5 and 0.6.
Asymptotic power law behavior with exponent dependent on $\gamma$ is observed for $\gamma\geqslant 1/2$. For $\gamma=1/2$ $Q_c-1\sim N^{-1}$, the decay being slower for greater $\gamma$.\label{Q_c-gamma}}
\end{figure}

A similar trend is observed for $q\geqslant 1$. Quotients $Q_0$ and $Q_c$ tend to 1 as $N$
increases for all values of $\gamma$, which, as
mentioned before, determines the growth of the interval where the $q-$Gaussian is evaluated. 
In the case of the Gaussian, i.e. $q=1$,
it is known that $\gamma=1/2$. Figure~\ref{Q_c-gamma} shows the decay of
the central quotient for $q=3/2$ and different values of $\gamma$. Apparently, $\gamma=1/2$
provides the appropriate growth of the interval for $q$--Gaussians with $q>1$ as well. For
$\gamma<1/2$, one observes the power-law behavior only over some range whereas for $\gamma>1/2$
the decay follows a power-law with an exponent larger than $-1$. The boundary ratio displays the
same dependence on $\gamma$. 

\begin{figure}[h!]
\centering\epsfig{file=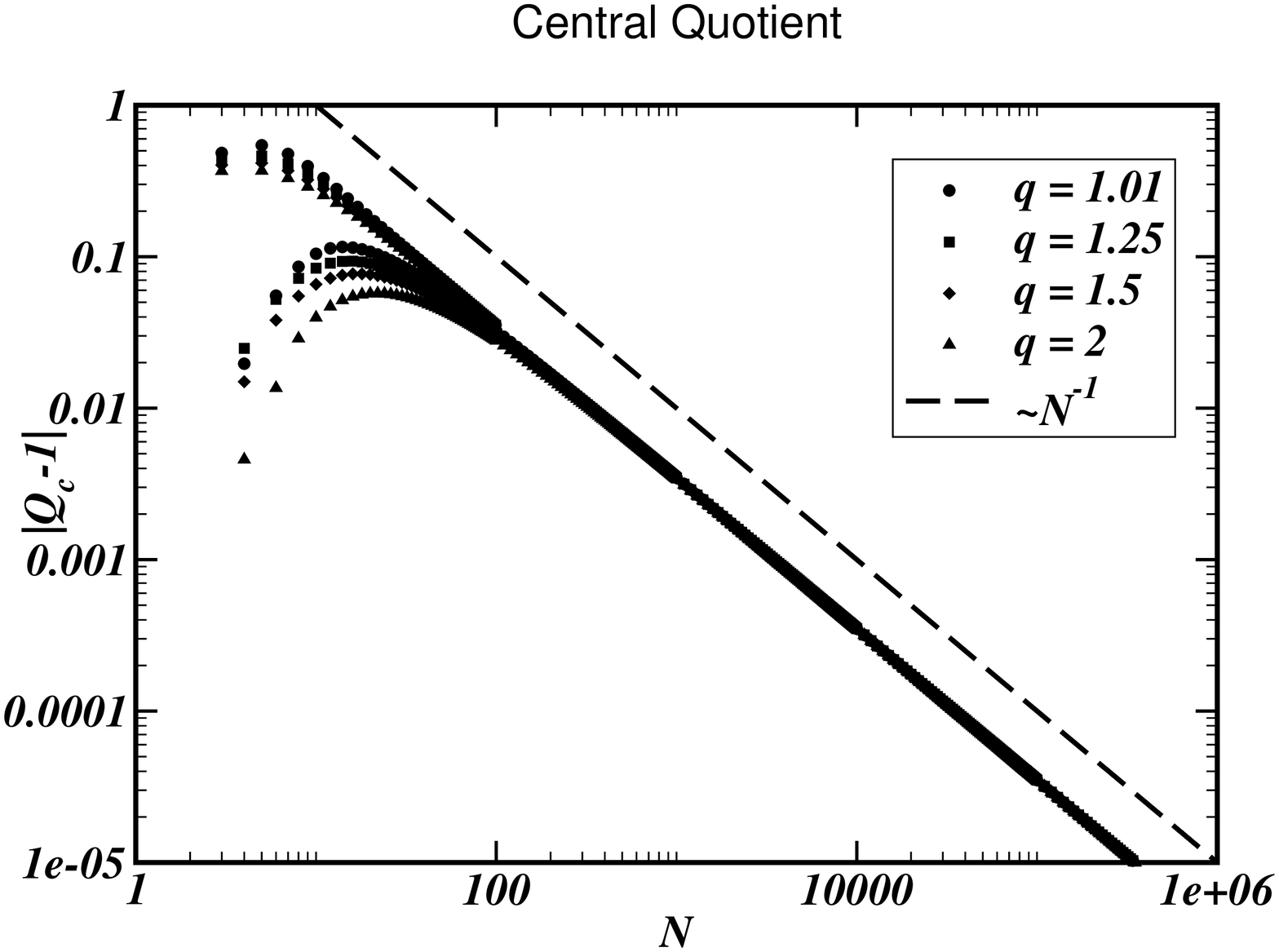, width=0.6\textwidth, clip=,angle=-90}
\centering\epsfig{file=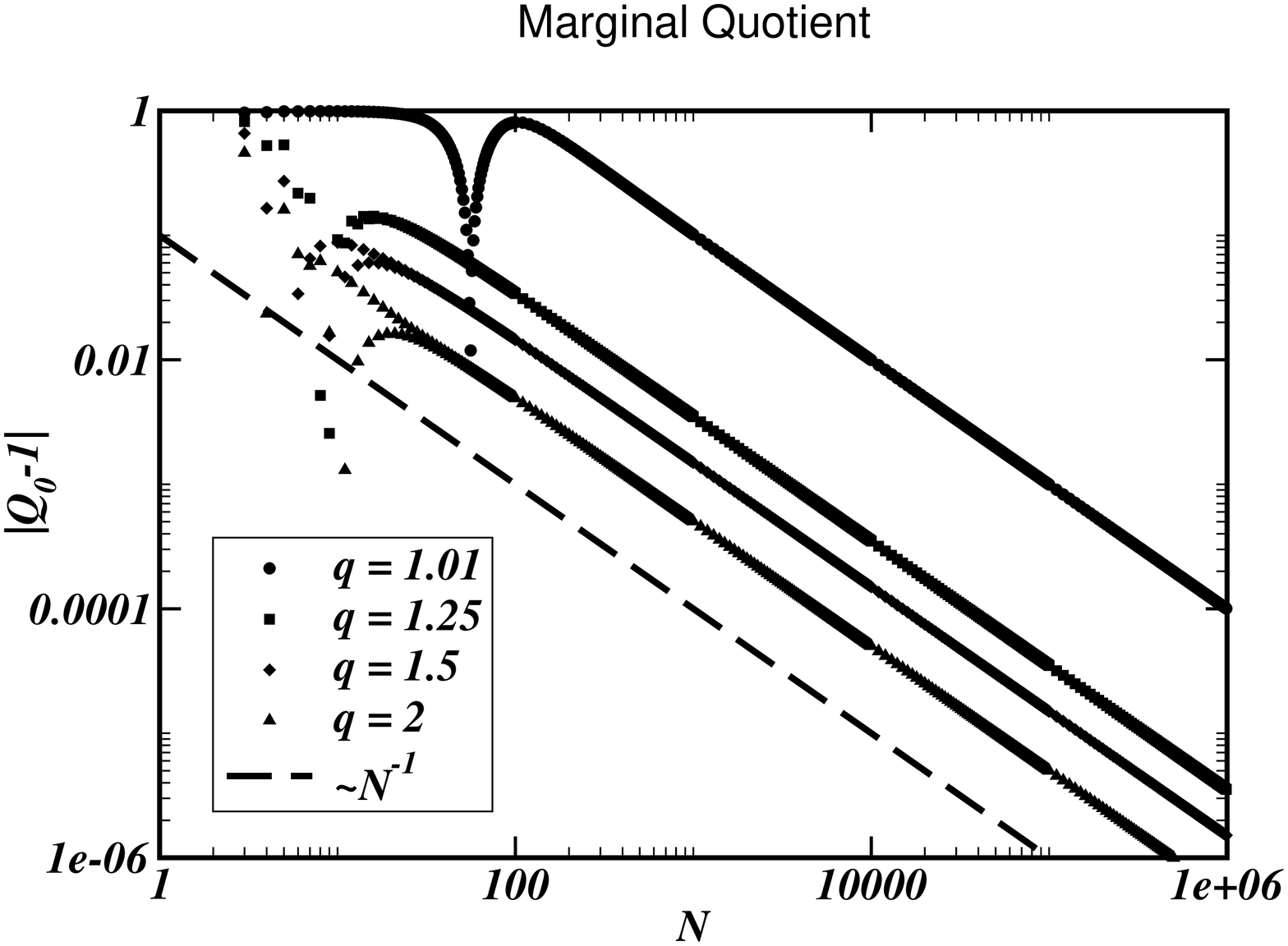, width=0.6\textwidth, clip=,angle=-90}
\caption{Central quotient (top) and boundary quotient (bottom) decay as a function of $N$ for discretized $q-$Gaussians with
  $q=1.01$, 1.25, 1.25 and 2, $\gamma=1/2$ and $\delta=2$. In all cases the trend $Q_0-1\sim Q_c-1\sim N^{-1}$ is
  observed. \label{Q_c-Q_0-gamma0.5}}
\end{figure}

For $\gamma=1/2$ we verify a $1/N$ power-law. Figure~\ref{Q_c-Q_0-gamma0.5} shows $Q_c-1$ and
$Q_0-1$ for typical values of $q$ and $\gamma=1/2$. 

\begin{figure}[h]
\begin{minipage}{.33\linewidth}
\epsfig{file=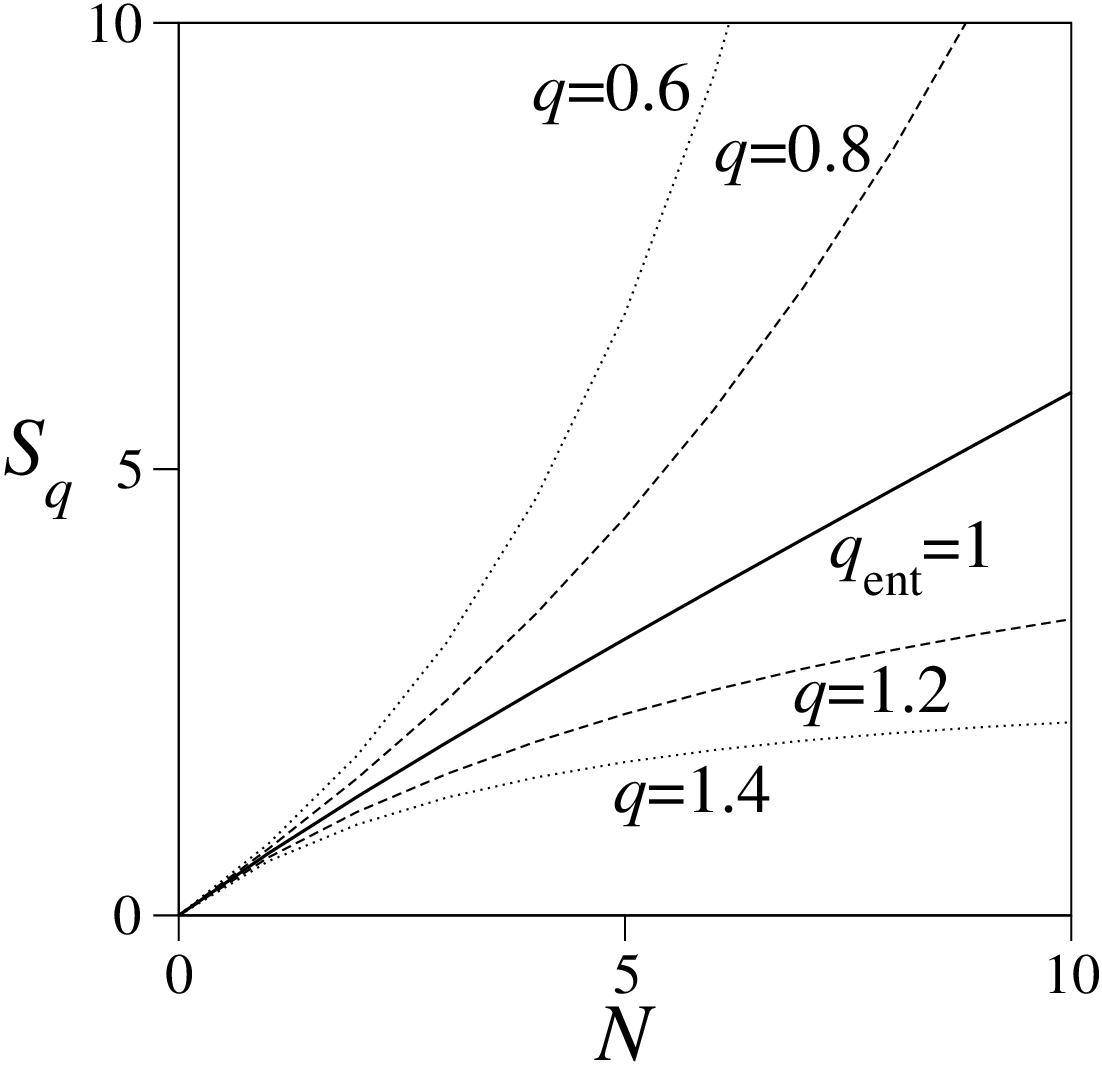, width=4.5cm, clip=}
\end{minipage}\hspace*{2mm}
\begin{minipage}{.33\linewidth}
\epsfig{file=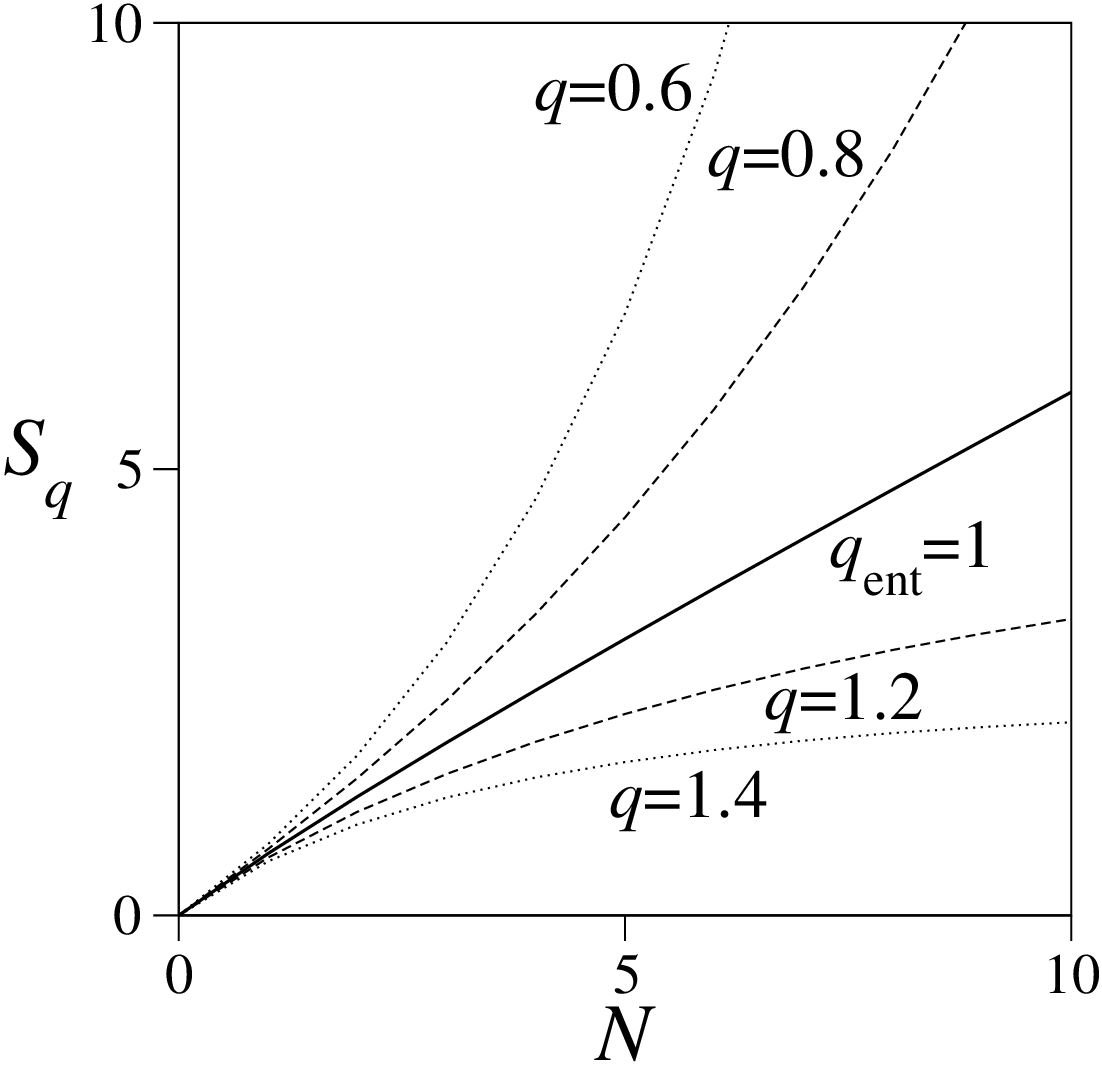, width=4.5cm, clip=}
\end{minipage}\hspace*{2mm}
\begin{minipage}{.33\linewidth}
\epsfig{file=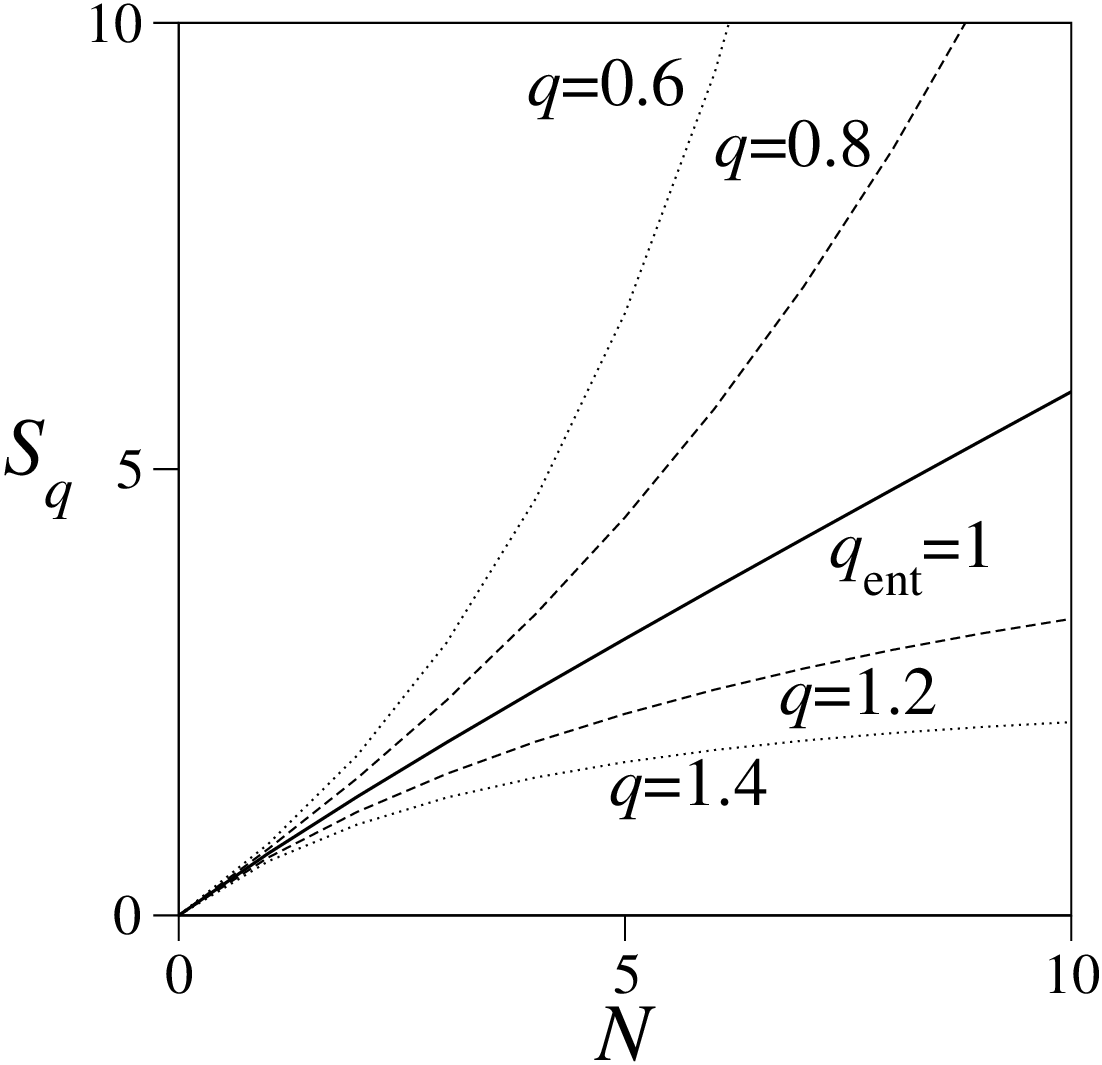, width=4.5cm, clip=}
\end{minipage}
\caption{ $q-$entropy as in FIG. \ref{S_q_q-G} for discretized $q-$Gaussians with $q=1$ (left), $q=3/2$ (center), and $q=2$
  (right). In all cases $q_\text{ent}=1$.\label{S_q>=1}}
\end{figure}

In what concerns the extensivity of the entropy $S_q$, the value of $q_\text{ent}$ remains 1 and is independent of
$\gamma$. Figure~\ref{S_q>=1} shows the $S_q(N)$ for typical values of $q\geqslant 1$.

\section{THIRD MODEL: ANOTHER FAMILY OF GENERALIZED TRIANGLES}
\label{sec:another}

As seen in Sections \ref{sec:family} and \ref{sec:discretized}, strictly as well as asymptotically scale-invariant probability models may lead to $q-$Gaussian limiting distributions. Nevertheless, as we already now~\cite{Hilhorst:07}, scale-invariance does not guarantee $q-$Gaussianity. In this Section, we present a last model to emphasize this point.

The following strictly scale-invariant triangle

\begin{tabular}{cccccccccccccccc}
$(N=0)\quad\quad$&&&&&&&&1&&&&&&&\\
$(N=1)\quad\quad$&&&&&&&$\dfrac{1}{2}$&&$\dfrac{1}{2}$&&&&&&\\
$(N=2)\quad\quad$&&&&&&$\dfrac{4}{14}$&&$\dfrac{3}{14}$&&$\dfrac{4}{14}$&&&&&\\
$(N=3)\quad\quad$&&&&&$\dfrac{5}{28}$&&$\dfrac{3}{28}$&&$\dfrac{3}{28}$&&$\dfrac{5}{28}$&&&&\\
$(N=4)\quad\quad$&&&&$\dfrac{7}{56}$&&$\dfrac{3}{56}$&&$\dfrac{3}{56}$&&$\dfrac{3}{56}$&&$\dfrac{7}{56}$&&&\\
$(N=5)\quad\quad$&&&$\dfrac{11}{112}$&&$\dfrac{3}{112}$&&$\dfrac{3}{112}$&&$\dfrac{3}{112}$&&$\dfrac{3}{112}$&&$\dfrac{11}{112}$&&\\
&&&&$\vdots$&&&&$\vdots$&&&&$\vdots$&\\\end{tabular}

with coefficients ($\chi = 3/7$) given by
\begin{equation}r_{N,n}=\left\{\begin{array}{lcl}\dfrac{1}{2}-\chi (1-2^{1-N});&&n=0,N\\\vspace*{-.2cm}\\
\chi 2^{1-N};&&n\neq 0,N\end{array}\right.\end{equation}
corresponds to a different way to introduce correlations in the system. In order to get nonnegative coefficients, the parameter $\chi$ is
kept within $[0,1/2]$.

The probabilities are given by
\begin{equation}
\label{eq:prob_Lconst}
p_{N,n} = \left(\dfrac{1}{2}-\chi \right) (\delta_{n,0}+\delta_{n,N})
+\binom{N}{n} \chi 2^{1-N} ~.
\end{equation}
The case of binary random variables ($\nu \to \infty$ of the triangles analyzed in Sec.~\ref{sec:family}) is reproduced here for $\chi=1/2$, hence 
  $p_{N,n} =\binom{N}{n} 2^{-N} ~$.
  
In order to calculate the limiting probability function, the CLT states that  the new variable  $x=(n-N/2)/(\sqrt{N}/2)$ provides a normal distribution in the limit $N\to\infty$ for the second term of Eq.~\eqref{eq:prob_Lconst}. In addition, two delta peaks appear after substitutions $\delta_{n,0}\to\delta(\frac{\sqrt N}{2}x+\frac{N}{2})=\frac{2}{\sqrt N}\delta(x+\sqrt N)$ and $\delta_{n-N,0}\to\delta(\frac{\sqrt N}{2}x-\frac{N}{2})=\frac{2}{\sqrt N}\delta(x-\sqrt N)$. Finally, by taking limits in Eq.~\eqref{eq:prob_Lconst}, we obtain     
\begin{equation}
\label{eq:Lconst_limit}
\mathcal{P}(x)=\lim\limits_{N\to \infty}\frac{\sqrt N}{2} p_{N,n}=2\chi \frac{1}{\sqrt{2\pi}} e^{-x^2/2}+\left(\frac{1}{2}-\chi\right)\lim_{N\to\infty}(\delta(x-\sqrt{N})+\delta(x+\sqrt{N}))
\end{equation}
which consists of a Gaussian distribution plus the additional contribution of the delta peaks corresponding to a concentration of probability on the two sides of the triangle.

As in the previous sections, the BG--entropy is extensive for this triangle as well. This may
be proved directly by inserting coefficients \eqref{eq:prob_Lconst} into
\begin{equation}
  \label{eq:entr_lastt}
  S_1 = - \sum \limits^{N}_{n=0} \binom{N}{n} r_{N,n} \ln{r_{N,n}}
\end{equation}
yielding
\begin{align}
  S_1 & =  -(1-\chi) \ln[(\dfrac{1}{2}-\chi)+\chi 2^{1-N}] - \chi 2^{1-N} \sum
  \limits^{N-1}_{n=1} \binom{N}{n} [\ln(\chi 2^{1-N})] \nonumber \\
  & = (\chi-1) \ln[(\dfrac{1}{2}-\chi)+\chi 2^{1-N}]-2\chi \left(\ln(2\chi)-2^{1-N} \ln(\chi
  2^{1-N})\right)+2\chi N \ln 2 \nonumber\\
& \propto N
  \label{eq:entr_lastt2}
\end{align}
for large $N$.

\section{CONCLUSIONS}
\label{conclusions}

A family of Leibnitz-like triangles, leading to $q$--Gaussians as limiting probability distribution functions
with $q\leqslant 1$, was introduced, where the limiting distribution could be exactly
calculated. These systems correspond to $N$ correlated binary random variables, the
index $q$ characterizing the strength of correlation. The case $q\to -\infty$ corresponds to very strongly
correlated variables giving a uniform limiting distribution. 

On the other hand, the coefficients of another type of triangles were constructed by
discretizing $q$--Gaussians. These triangles, having now by construction $q$--Gaussians with $q<1$ as
limiting probability functions, showed a behavior with depends on the specific discretization of
the support interval. Except for one particular case, the
Leibnitz rule, related to system size scale invariance of the probabilities, is only
asymptotically satisfied. The system approaches scale invariance with a $1/N^2$ power-law for large $N$, except for the boundary coefficients where the convergence to scale
invariance is much slower, of the type $1/N$. The $1/N^2$ law makes a crossover into a $1/N$ one over the entire triangle when considering $q$--Gaussians with $q\geqslant 1$. 

Finally, another family of strictly scale-invariant triangles with a rather strange limiting distribution function was introduced. In the limit $N\to \infty$, the triangles yield a Gaussian distribution together with two delta peaks centered at points going to infinity. 

The BG--entropy remains extensive for all three types of triangles, equally to previously
studied Leibnitz--like triangles~\cite{MoyanoGellmann:2006}. This may be the result of the
simplicity of the models presented in this paper. More sophisticated models, as for instance 
the Hamiltonian mean field model (see for instance ref.~\cite{Pluchino:07,Pluchino:08}), appear to approach a
$q$--Gaussian characterized by a non-equilibrium stationary state with the $q$--entropy possibly being
extensive for $q\neq1$. However, in the present effort we are here not particularly interested in the general relation
between the extensivity of the entropy and stationary-state probability distributions, but we rather searched to find out which kind of
correlation between the microscopic events of a system leads to $q$--Gaussians as limiting distributions (possibly, as attractors).

The Leibnitz rule provides a simple tool to study models composed 
of correlated binary random variables, and enabled the exact calculation of their limiting functions. 
As already addressed in ~\cite{Hilhorst:07}, this rule cannot be uniquely related
to nonextensive thermostatistics. Indeed, Leibnitz-like triangles exist which precisely lead to
$q$--Gaussians (as shown in the present paper) as well as to other limiting probability functions (as shown in \cite{Hilhorst:07}, and also here). Additionally, the present second
family of triangles (with asymptotic but not strict scale invariance) also tended to a $q$--Gaussian. The scenario which emerges is that
asymptotic validity of the Leibnitz rule might represent a necessary but surely not
sufficient condition for the system to tend to $q$--Gaussians as limiting distributions when $N\to\infty$. 

The fact that different implementations of correlations between the variables of a system can
lead to the same function, --- $q$--Gaussians in the present case ---, can be seen as a hint for these
functions being attractors for a variety of different systems, and so supports the
demand of generality of the $q$-generalized central limit theorem presented in
~\cite{Tsallis:05,Umarov:08,Umarov:06a,Umarov:07}. However, to assure the applicability of this
central limit theorem, the stability of the $q$--Gaussians as limiting functions of the 
 systems presented here needs to be proved, either by establishing that the correlations correspond to $q$-independence, or  by introducing, for example, weak perturbations. 

\begin{acknowledgments}
We acknowledge fruitful remarks by H.J. Hilhorst and S. Umarov, as well as partial financial support by CNPq and FAPERJ (Brazilian Agencies) and DGU-MEC (Spanish Ministry of Education) through Projects MOSAICO and PHB2007-0095-PC. 
\end{acknowledgments}


\end{document}